\documentclass[11pt,a4paper]{article}

\usepackage{endnotes}
\usepackage{ulem}
\usepackage{color}
\let\footnote=\endnote

\newcommand{\vct}[1]{\boldsymbol{#1}}
\newcommand{\mtrix}[3]{\langle \,#1\,|\,#2\,|\,#3\,\rangle}
\newcommand{\avr}[1]{\langle \,#1\,\rangle}
\usepackage{amsmath, amssymb}
\usepackage{bm}

\usepackage[dvipdfmx]{graphicx}
\usepackage{amsmath, amssymb}

\begin{document}

\begin{center}
{\Large
Moments of the charge distribution observed through
electron scattering in $^3$H and $^3$He
}

\vspace{5mm} 
\noindent
  
Emiko Hiyama$^1$ and Toshio Suzuki$^2$
 
$^1$Department of Physics,
Tohoku University Sendai 980-8578, Japan, and\\ 
RIKEN Nishina Center, 2-1, Hirosawa, Wako, 351-0198, Japan\\
 $^2$Research Center for Accelerator and Radioisotope Science,
Tohoku University \\Sendai 982-0826, Japan

\end{center}

\vspace{3mm}
\noindent
kt.suzuki2th@gmail.com

\vspace{5mm}
\noindent
Abstract :
The moments of the charge distributions obtained by the sum-of-Gaussians(SOG)-analysis
of electron-scattering data are examined in $^3$H and $^3$He, together with those by
the Fourier-Bessel(FB)-one.
The SOG- and FB-methods reproduce well the experimental form-factors available at present,
but provide different charge-distributions from each other.
As a result, they do not yield the same values of the moments 
of the charge distribution,
although their analysis are called `model-independent'.  
The moments are sensitive to the tail of the charge distribution.
The present experimental data are not enough for SOG- and FB-analyses
to determine the reasonable shape of the tails, in a quantum mechanical point of view.
The new accurate experimental data at small momentum transfer squared
less than $0.1$ fm$^{-2}$ are desired.


\section{Introduction}\label{intro}

Electron-scattering experiments have provided continuously us
with fundamental information of nuclear structure
for long time\cite{bd, deforest, bm,vries}.
Since electromagnetic interaction is well
understood theoretically\cite{bd,deforest},
the reaction mechanism
is almost completely separated from assumptions on the nuclear structure
which is dominated by strong interaction.
Hence, the observed form factors of
$^3$H and $^3$He, as mirror nuclei with the least number of constituent nucleons,
have played a role of touchstones for understanding structure of more complex nuclei
from the beginning of electron-scattering history\cite{schiff,coll}.
Around in 1980, in addition to the nucleonic degrees of freedom,
the mesonic effects on the form factors of the few-body systems were extensively
studied along with new experiment\cite{beck,jus, ott,am}
and development of the calculation-method
based on realistic nuclear forces\cite{sch,kat,str,riska,sickr}.
The meson exchange current(MEC) has been shown to be necessary for understanding
the experimental form-factors\cite{am}.
The form factor provides the charge density distribution, $\rho_c(r)$, in the nucleus.
The gross profiles of $\rho_c(r)$ in the three-nucleon systems
were thus established about 30 years ago, using the world data at that time\cite{am}.


The $n$th moment of $\rho_c(r)$ is defined by 
$\avr{r^n}_c=\int dr^3 r^{n} \rho_c(r)/Z$,
$Z$ denoting the number of the protons in the nucleus.
The second moment($n=2$)
is called the mean square radius(msr)\footnote{The abbreviation
of the `rms'(root mean square)-radius is frequently used in the literature,
but it is convenient for the present purpose to employ `msr' for
the mean square radius, because electron scattering observes the value of the msr
($\avr{r^2}_c$), together with the higher moments, $\avr{r^{2n}}_c, (n=2, 3, \cdots)$,
rather than the square root of the msr.},
and is one of the key quantities in studying nuclear structure\cite{bm}.
The msrs of $^3$H and $^3$He, as fundamental systems of the complex nuclei, should be
determined experimentally as accurately as possible.
In fact, however, we can not say that the values of the msrs are accurate enough for
recent requirements in nuclear physics, in spite of the fact that
the shape of $\rho_c(r)$ is explored with the use of the `world data'.


There are two ways at present to deduce the shape of $\rho_c(r)$ 
from the experimental cross sections of electron scattering.
They are called the sum-of-Gaussians(SOG)\cite{sog}- and
the Fourier-Bessel(FB)\cite{fl,fb}-method,
which are referred frequently as a `model-independent' way\cite{vries}.
If experimental cross sections were available for the values of $q$ from $0$ to $\infty$,
$q$ being the momentum transfer from the
electron to the nucleus, both ways would yield $\rho_c(r)$ `model-independently',
which reproduces cross sections, according the phase-shift analysis
of the experiments,
and then the values of the msrs would uniquely be determined.
The $q$-region in the experiments performed so far, however,
is limited even in `world data' for $^3$H and $^3$H \cite{am}. 
Hence, strictly speaking,
the obtained $\rho_c(r)$, and therefore the value of the msr depends on the analysis-methods.
Indeed, Ref.\cite{vries} lists the two different values of the square root of the msr  
by the SOG- and the FB-method. For $^3$H, the former yields $1.76(4)$ and the latter
$1.68(3)$, while for $^3$He, the former $1.844(45)$
and the latter $1.877(19)$ in units of fm.
Ref.\cite{am} provides $1.753(86)$ for $^3$H and $1.959(30)$ fm for $^3$He
by the SOG-analysis with the use of the
 `world data', adding new experimental results after Ref.\cite{vries}.
Recent papers\cite{din, atom} have referred a different value(1.973(14) fm)
for $^3$He obtained by the SOG-method in Ref.\cite{sickz}.
In this way, the values of the msrs are strongly dependent on
the methods such as SOG and FB.
The SOG- and FB-methods were successful in studying the gross shape of $\rho_c(r)$\cite{am},
but not enough for providing accurate values of the msrs.

The purpose of the present paper is to explore in detail
how the values of the moments depend on the analysis-methods of electron-scattering data
in $^3$H and $^3$He.
Such an investigation is desirable at present, because
recent detailed calculations on few-body systems require
the accurate experimental values of the msr to be reproduced\cite{din,sickhe4},
while those aiming {\it ab initio} calculations employ the value of the msr of 
a few-body system as a reference in constructing nuclear interactions\cite{hagen}.
Moreover, Refs.\cite{ks1,kss,ts} have shown that
if the values of the higher moments, $\avr{r^n}_c(n\ge 4)$, are determined accurately,
both the proton- and neutron-distributions are investigated using the same experimental
results obtained through electromagnetic interaction. 
The trinucleon-systems, $^3$H and $^3$He, are mirror nuclei,
and at the same time, the most extremely neutron-rich and proton-rich among nuclei
explored experimentally so far in electron scattering.
We note also that recent experiment\cite{atom} on the laser spectroscopy
of muonic $^3$He-ions
provides the square root of the msr to be 1.97007(94) fm
with remarkable small errors,
against $1.959(30)$ in electron scattering\cite{am}.
The corrections for the msr from the muonic atom depends on the values of
the Zemach moments
which are related to the higher moments determined via electron scattering\cite{din,sickz}.

In the next section, descriptions of $\avr{r^n}_c$
will be reviewed, according to Refs.\cite{ks1,kss}.
\S \ref{av} will  briefly mention the calculations on the density
distributions of the three-body systems. 
In \S \ref{form}, the relationship between the form factors
and $\rho_c(r)$ will be discussed.
It will be shown that
the SOG- and FB-methods reproduce well the experimental form factors,
but yield the different charge distributions from each other,
because the experimental cross sections availabe at present are limited in the
region of $q$, as $q_{\rm min}\le q \le q_{\rm max}$. 
In particular, both methods fail to describe the tails of $\rho_c(r)$
expected in quantum mechanical calculations, which are sensitive to the
form factors in the region of $0<q\le q_{\rm min}$.
The MEC is necessary for understanding
the form factors of $^3$H and $^3$He\cite{am}, but is almost independent of the values
of $\avr{r^n}_c$. As a result, the moments of the point nucleon distributions are
explored using the charge distributions without the MEC-contributions.
In \S \ref{moment}, the values of $\avr{r^n}_c$ will be calculated, together with
those of the protons and the neutrons in $^3$H and $^3$He.
It will be shown that
the SOG\cite{sog}- and the FB\cite{fl,fb}-method provide a different `experimental
values' of $\avr{r^n}_c$ from each other, because of their different tails of $\rho_c(r)$.
In \S \ref{dis} are discussed some ways, instead of the SOG- and the FB-method,
to derive the values of $\avr{r^n}_c$
from experiment model-independently
as possible as we can. The proposed ways
require new accurate experiments at low $q$ as $0<q^2<0.1$ fm$^{-2}$. 
The final section, \S \ref{sum}, is devoted to a brief summary.

\section{Definition of the charge distribution and its moments}\label{mom}

In order to describe the charge density with the relativistic corrections,
we use the Foldy-Wouthuysen(FW)-method\cite{bd}
which makes the unitary transformation of the four-component
framework to the two-component one.
Because we do not know the original relativistic Hamiltonian, however,
we employ the Dirac equation with electromagnetic field,
in the same way as the previous authors in \cite{bertozzi,macvoy,nishizaki,kss}.

When writing the form factor in the plane-wave
Born approximation(PWBA) in elastic electron scattering as
\begin{equation}
 F_c(q)=\frac{1}{Z}\mtrix{0}{\hat{\rho}(q)}{0}\label{nfm}
\end{equation}
with the matrix element provided by the
the ground-state wave function in the two component framework,
the charge operator, $\hat{\rho}(q)$, given by the FW-transformation
up to order $1/M^2$, $M$ being the nucleon mass,
is written as\cite{kss} 
\begin{align}
\hat{\rho}(q) = \sum_{k=1}^Ae^{i\vct{q}\cdot\vct{r_k}}\Bigl( D_{1k}(q^2)
 + iD_{2k}(q^2)\vct{q}\!\cdot\!(\,\vct{p}_k\times\vct{\sigma}_k\,)\Bigr).
\label{fw}
\end{align}
Here, $Z$ and $A$ denote the numbers of the protons and nucleons of the system, respectively.
In the above equation, the first and the second term in the right-hand side
are called the Darwin-Foldy(DF) and the spin-orbit terms, respectively\cite{macvoy,ks0}.
They are defined by
\begin{align}
 D_{1k}(q^2)&= F_{1k}(q^2)-\frac{q^2}{2}D_{2k}(q^2), \label{d0m}\\[4pt]
 D_{2k}(q^2)&= \frac{1}{4M^{2}}\left(F_{1k}(q^2)+
 2\mu_k F_{2k}\right),\label{d2m}
\end{align}
with the Dirac form factor, $F_{1\tau}(q^2)$, related to the Sachs, $G_{E\tau}(q^2)$,
and Pauli, $F_{2\tau}(q^2)$, form factor as\cite{bd}
\begin{equation}
F_{1\tau}(q^2)=G_{E\tau}(q^2)+\mu_\tau q^2F_{2\tau}(q^2)/(4M^2),\label{sachs}
\end{equation}
where $\tau=p(n)$ indicates $k$ being the proton(neutron). 
Moreover, $\mu_\tau$ denots the anomalous
magnetic moment to be $\mu_\tau=1.793$ for $p$ and $-1.913$ for $n$. 
The momentum-transfer dependence of the nucleon form factors
is still under discussions theoretically\cite{sachs,licht,gmiller}.
Experimentally also there are various functional forms to fit
the electron scattering data at present\cite{kelly,kelly2}.
In the present paper,
the following Sachs and Pauli form factors 
will be employed\cite{kss,ts},
\begin{align}
G_{Ep}(q^2)&= \frac{1}{(1+r_p^2q^2/12)^2},
 \qquad F_{2p}=\frac{G_{Ep}(q^2)}
 {1+q^2/4M^2}, \label{expff}\\[4pt]
G_{En}(q^2)&= \frac{1}{(1+r_+^2q^2/12)^2}- \frac{1}{(1+r_-^2q^2/12)^2},\qquad
 F_{2n}=\frac{G_{Ep}(q^2)-G_{En}(q^2)/\mu_n}{1+q^2/4M^2},\nonumber
\end{align}
with the values used in Ref.\cite{rhoro}
\begin{equation}
  r_p=0.877\, \textrm{fm}, \qquad r_{\pm}^2=(0.830)^2
  \mp0.058 \, \textrm{fm}^2.\label{newr}
\end{equation}
In Ref.\cite{rhoro}, $G_{En}(q^2)$ is given by the form
\begin{equation}
 G_{En}(q^2)=-\frac{r_n^2q^2/6}{1+q^2/M^2}\frac{1}{(1+r_p^2q^2/12)^2},
\end{equation}
with $r_n^2=-0.116\, {\rm fm}^2$. This is numerically almost equal to
$G_{En}(q^2)$ in Eq.(\ref{expff}) with the values of Eq.(\ref{newr}),
and the values of the first and the second derivative of these form factors
are taken to be equal to each other at $q^2=0$.
There are still discussions on the values of $r_p$ and $ r_{\pm}^2$
themselves\cite{sick,data1,data2,xiong}.
The value $r_p=0.877$ fm is almost equal to the upper bound
$r_p=0.887$ fm of the proton size at present\cite{sick}.
The uncertainty of the nucleon-size, however, does not change our main conclusions
with respect to the SOG- and FB-methods in the present paper.
The charge density, $\rho_c(r)$, with the relativistic corrections
is given by the Fourier transform of Eq.(\ref{nfm}) as
\begin{equation}
 \rho_c(r) = Z \int \frac{d^3q}{(2\pi)^3}\exp(-i\vct{q}\!\cdot\!\vct{r})
 F_c(q).\label{rcd}
\end{equation}

The mean $2n$th-order moment, $\avr{r^{2n}}_c$,
of $\rho_c(r)$ is given for $n=1,2,3 \cdots$ by
\begin{equation}
\avr{r^{2n}}_c =\sum_\tau \avr{r^{2n}}_{c\tau},
\qquad
\avr{r^{2n}}_{c\tau}=\frac{1}{Z}\int d^3r\, r^{2n}\rho_{c\tau}(r),\label{nth1}
\end{equation}
where $\rho_{c\tau}(r)$ stands for the proton(neutron) charge distribution for $\tau=p(n)$,
as $\rho_c(r)=\rho_{cp}(r)+\rho_{cn}(r)$.
In calculating $\avr{r^{2n}}_c$, it is convenient to use the following
identity instead of the above equation itself \cite{ks1},
\begin{equation}
\avr{r^{2n}}_{c\tau}
 =(-\vct{\nabla}_q^2)^n F_{c\tau}(q)|_{q=0},
\end{equation}
where $F_{c\tau}(q)$ for $\tau=p(n)$ denotes the part of Eq.(\ref{nfm})
due to the protons(neutrons) as
\begin{equation}
F_{c\tau}(q)=\frac{1}{Z}\int dr^3\exp(i\vct{q}\!\cdot\!\vct{r}) \rho_{c\tau}(r).\label{npfm}
\end{equation}

In the present discussions on $^3$H and $^{3}$He, contributions of the
relativistic corrections to the form factors are negligible\cite{sch},
but the DF-terms will be kept in some calculations of the moments,
while the spin-orbit term is neglected, because it is expected
not to be necessary for the present purpose\cite{kss,ts}.
In denoting the $2n$th moment of the {\it point} proton and neutron
distributions by $\avr{r^{2n}}_\tau \,\, (\tau=p, n)$,  
the explicit expressions for the moments$(n\le 4)$ of the charge distribution,
with the DF-corrections,
are written as follows.
\begin{eqnarray}
\avr{r^2}_c&=&\avr{r^2}_p+r_p^2+\frac{N}{Z}r^2_n+\frac{3}{4M^2_p},
 \quad r^2_n=(r^2_+-r^2_-),\label{msr}\label{2nd}\\
\avr{r^4}_c&=&\avr{r^4}_{cp}-\avr{r^4}_{cn},  \\
&&\avr{r^4}_{cp}=\left(\avr{r^4}_p++\frac{5}{2M^2_p}\avr{r^2}_p\right)+\frac{10}{3}r^2_p(\avr{r^2}_p
 +\frac{3}{4M^2_p})+\frac{5}{2}r^4_p,
 \nonumber\\
&&\avr{r^4}_{cn}=-\frac{N}{Z}\left[\frac{10}{3}(r^2_+-r^2_-)(\avr{r^2}_n+\frac{3}{4M^2_n})
 +\frac{5}{2}(r^4_+-r^4_-)\right],
 \nonumber\\
 \avr{r^6}_c&=&\avr{r^6}_{cp}-\avr{r^6}_{cn},\\
&&\avr{r^6}_{cp}=\left(\avr{r^6}_p+\frac{21}{4M^2_p}\avr{r^4}_p\right)
 +7r^2_p\left(\avr{r^4}_p+\frac{5}{2M^2_p}
\avr{r^2}_p\right) \nonumber\\
 &&\qquad\qquad\qquad+\frac{35}{2}r^4_p\left(\avr{r^2}_p
+\frac{3}{4M^2_p}\right)+\frac{35}{3}r^6_p\label{hp},\nonumber\\
&&\avr{r^6}_{cn}
 =-\frac{N}{Z}\left[7(r^2_+-r^2_-)\left(\avr{r^4}_n+\frac{5}{2M_n^2}
\avr{r^2}_n\right)+\frac{35}{2}(r^4_+-r^4_-)\left(\avr{r^2}_n+\frac{3}{4M^2_n}\right) \right. \nonumber\\
&&\left.\qquad\qquad\qquad+\frac{35}{3}(r^6_+-r^6_-)\right],\label{hn}\nonumber
\end{eqnarray}
  \begin{eqnarray}
\avr{r^8}_c&=&\avr{r^8}_{cp}-\avr{r^8}_{cn},\label{8th}\\
&&\avr{r^8}_{cp}=\left(\avr{r^8}_p+\frac{9}{M_p^2}\avr{r^6}_p\right)
 +12r^2_p\left(\avr{r^6}_p+\frac{21}{4M_p^2}\avr{r^4}_p\right)
\nonumber\\
&&\qquad\qquad\qquad   +63r^4_p\left(\avr{r^4}_p+\frac{5}{2M_p^2}\avr{r^2}_p\right)
+140r^6_p\left(\avr{r^2}_p+\frac{3}{4M_p^2}\right)
+\frac{175}{2}r^8_p\,,\nonumber\\
&&\avr{r^8}_{cn}=-\frac{N}{Z}\left[12(r^2_+-r^2_-)\left(\avr{r^6}_n+\frac{21}{4M_n^2}\avr{r^4}_n\right)
		       +63(r^4_+-r^4_-)\left(\avr{r^4}_n+\frac{5}{2M_n^2}\avr{r^2}_n\right)\right.\nonumber\\
&&\left.\qquad\qquad\qquad +140(r^6_+-r^6_-)
		       \left(\avr{r^2}_n+\frac{3}{4M_n^2}\right)+\frac{175}{2}(r^8_+-r^8_-)\right].\nonumber
\end{eqnarray}
Here, $M_{p(n)}$ denotes the proton(neutron) mass, whose value is taken to be equal to $M=939$ MeV
in the following calculations.
As seen in the above equations, contributions of the neutron-distribution appear
in the $2n(\ge 4)$th moments of the charge distributions.

\section{ Density distributions of $^3$H and $^3$He} \label{av}

\begin{figure}[htb]
\begin{center}
\includegraphics[scale=0.5]{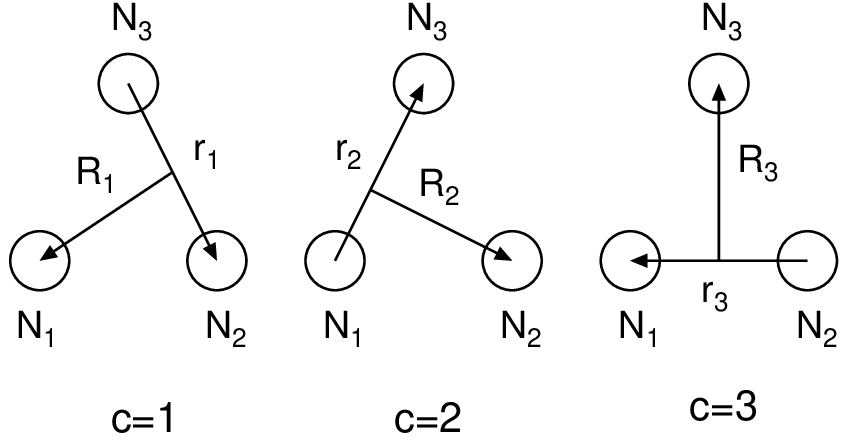}
\end{center}
\caption{Jacobian coordinates for three nucleons.}
\label{jacobi}
\end{figure}

The density distributions of the three-body systems are calculated
by using the AV8' $+3N$ potential\cite{He4}.
The Argonne V8'-potential(AV8') \cite{AV8}
is derived from the realistic
AV18\cite{AV18} by neglecting
the charge dependence and the terms
proportional to
$L^2$ and $(L \cdot S)^2$. 
The use of the AV8' fails
to reproduce the experimental values of 
the binding energies of
$^3$H, $^3$He and $^4$He,
which are important for the present discussions
on the density distributions.Thus,
we employ the AV8'$+3N$ potential which adds to AV8'
a phenomenological
$3N$ force given by a sum of two Gaussian terms\cite{He4}:
\begin{equation}
V_{ijk}^{3N}=\sum^2_{n=1}W_n {\rm exp}[(-r^2_{ij}+r^2_{jk}
+r^2_{ki})/b^2_n]{\cal{P}}_{ijk}
\end{equation}
with 
\begin{eqnarray}
W_1 =-2.04\, {\rm MeV}, \quad b_1=4.0\, {\rm fm}, \nonumber \\
W_2=35.0\, {\rm MeV}, \quad b_2=0.75\, {\rm fm}.
\end{eqnarray}
The above parameters give the following
binding energies of
$^3$H and $^3$He: 8.41 MeV and 7.74 MeV, respectively,
which are in good agreement with
data: 8.48 MeV for $^3$H and 7.72 MeV for $^3$He.
The parameters for the phenomenological potential are also
used in the calculations of the transition form factor for
$^4{\rm He}(e,e')^4 {\rm He}(0^+_2)$ \cite{He4}.

To solve three nucleon systems such as $^3$H and $^3$He,
we use Gaussian Expansion Method (GEM)\cite{GEM}.
The Schroedinger equation is written as
\begin{equation}
[T+V(r_1)+V(r_2)+V(r_3)-E]\Psi_{JM,TT_z}=0,
\end{equation}
where $T=$ is the kinetic energy operator and $V(r_n)$ is the
potential between two nucleons shown in Fig \ref{jacobi},
The three-body total wave function $\Psi_{JM,TT_z}$
is described as a sum of three rearrangement channels
$c=1-3$ shown in Fig. \ref{jacobi}.
\begin{equation}
\Psi_{JM, TT_z}=\sum^3_{c=1} \Phi^{(c)}_{JMTT_z}(\bf{r_c,R_c}). \label{total}
\end{equation}
The $J$ and $M$ are total spin and its $z$-component and $T$ and $T_z$
are the total isospin and its $z$-component.
We consider the case of $J=T=1/2$ and $T_z=1/2$ for $^3$He and $-1/2$
for $^3$H.
The Jacobian coordinates, ${\bf r_c}$ and ${\bf R_c}$,
imply 
${\bf r_k}={\bf x_i}-{\bf x_j}$ and ${\bf R_k}={\bf x_k}-({\bf x_i}+{\bf x_j})/2$
for the cyclic permutations of three particles, $(i,j,k)$.
Each amplitude is given by
\begin{equation}
\Phi^{(c)}_{JMTT_z}({\bf r_c,R_c})=[\eta_t(jk)\eta_{1/2}(i)]_{TT_z}
\left[[\phi_{n\ell}({\bf r_c})\phi_{NL}({\bf R_c})]_\Lambda [\chi_s(jk)\chi_{1/2}(i)]_\Sigma]\right]_{JMTT_z}.\label{c}
\end{equation}
Here,
$\chi$ and $\eta$ are the spin and isospin functions, respectively,
and the basis functions, $\phi({\bf r_c})$ and  $\phi({\bf R_c})$,
are described by the Gaussian function and spherical harmonics:
\begin{eqnarray}
\phi_{n\ell m({\bf r_c})}&=&\phi_{n\ell}(r_c) Y_{\ell m}(\hat{\bf r_c}),
\quad \phi_{n\ell}(r_c) =N_{n\ell}r^{\ell} e^{-\nu r^2}, \nonumber \\
\phi_{N L M({\bf R_c})}&=&\phi_{N L}(R_c) Y_{L M}(\hat{\bf R_c}),
\quad \phi_{N L}(R_c) =N_{NL}r^{L} e^{-\lambda R^2}, 
\end{eqnarray}
 where
$N_{n\ell}$ and $N_{NL}$  are the normalization constants, and
the Gaussian range parameters, $\nu$ and $\lambda$, are
 taken as geometric progression.
The detail for GEM is written in Ref.\cite{GEM}.

The point neutron and proton
densities, $\rho_\tau(r)$, are calculated as a function of $r$
from the center-of-mass of the three-body systems,
$^3$H and $^3$He\cite{He4}.

\section{ Charge distribution and form factor}\label{form}

\begin{figure}[ht]
\begin{minipage}[t]{7.2cm}
\includegraphics[scale=0.9]{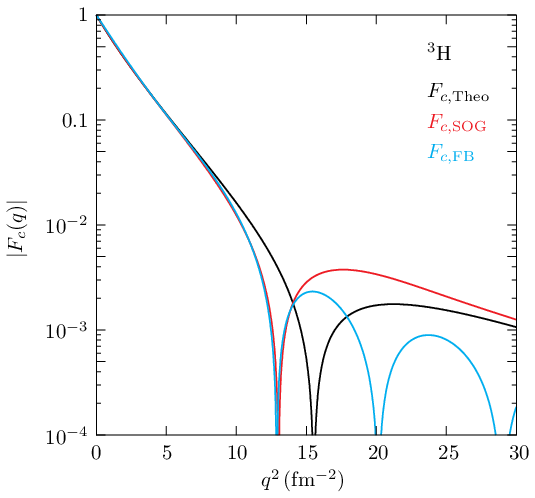}%
\hspace*{-1.5cm}(a)
\end{minipage}\hspace{0.8cm}%
\begin{minipage}[t]{7.2cm}
\includegraphics[scale=0.9]{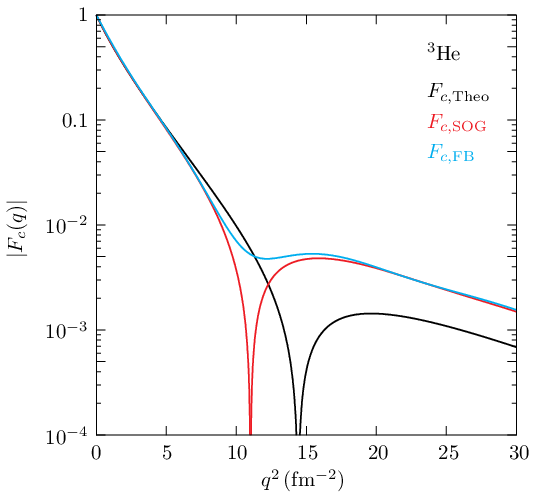}%
\hspace*{-1.5cm}(b)

\end{minipage}
\caption{
The form factors of $^3$H(a) and $^3$He(b)
as a function of the
 momentum transfer squared, $q^2$, in units of fm$^{-2}$.
The red curves are obtained by the sum-of-Gaussians(SOG)-analysis of
the experimental data\cite{am},
the blue ones by the Fourier-Bessel(FB)-analysis\cite{vries},
and the black ones by the present
calculations with AV8 in PWBA.
For details, see the text.
}
\label{fig_form}
\end{figure}

\begin{figure}[ht]
\begin{minipage}[t]{7.2cm}
\includegraphics[scale=0.9]{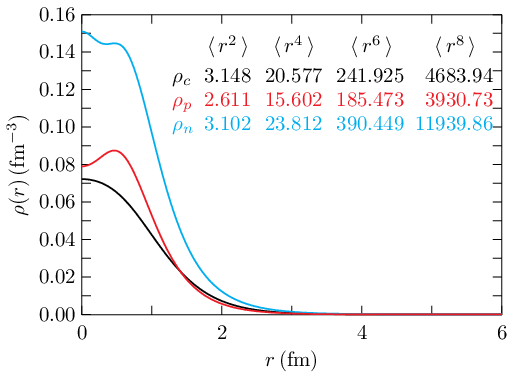}%
\hspace*{-1.5cm}(a)
\end{minipage}\hspace{0.8cm}%
\begin{minipage}[t]{7.2cm}
\includegraphics[scale=0.9]{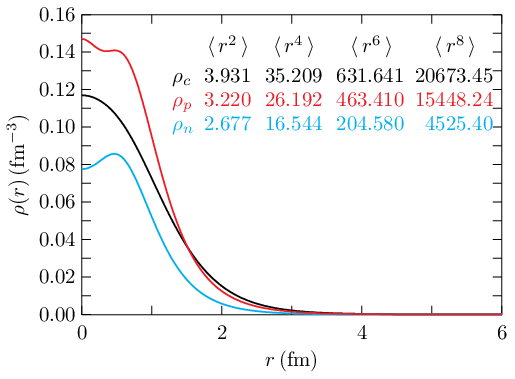}%
\hspace*{-1.5cm}(b)

\end{minipage}
\caption{
The point proton(red), point neutron(blue), and total charge (black) distributions
 in $^{3}$H on the left-hand side and in $^3$He on the right-hand side,
 as a function of $r$.
 The values of the corresponding $n$th moment, $\avr{r^n}$, are
also given in the each figure in units of fm$^n$. 
For details, see the text.
}
\label{point}
\end{figure}

The FB- and the SOG-method determine the charge density so as to reproduce
the experimental cross-section in the phase-shift analysis\cite{vries}.
If the distortion effects on the electron wave are taken into account
appropriately\cite{am,beck,ott},
the PWBA form factors\cite{deforest} in Eq.(\ref{nfm}),
which are the Fourier transform of $\rho_c(r)$ in Eq.(\ref{rcd}),
are convenient for theoretical investigation, 
because they reflect nuclear structure much more transparently.
Ref.\cite{am} provided the PWBA form factors inferred from the phase-shift
analysis of the world data by the SOG-method.
The data were in the region of $q^2$
between $0.09\le q^2 \le 30$ fm$^{-2}$ for $^3$H, 
and between $0.032\le q^2\le 30$ and $^3$He\cite{am}.
We will use those PWBA form factors in Fig. 9 of Ref.\cite{am},
as the experimental data, throughout this paper,
when we compare our results calculated by Eq.(\ref{nfm}) with experiment.

Figure 9 of Ref.\cite{am} is almost reproduced in Fig. \ref{fig_form}
by the present calculations.
The red curves, $F_{c,{\rm SOG}}$, which are understood as the experimental form factors,
are obtained by the Fourier transform of the SOG charge densities given in Ref.\cite{am}.
Although the distortion effects are not taken into account explicitly,
the minimum-position and the maximum-value of the second peak in the form factor,
particularly emphasized in Ref.\cite{am}, are almost the same
as those of Fig. 9 for both $^3$H and $^3$He.
The black curves, $F_{c,{\rm Theo}}$, show the form factors in Eq.(\ref{nfm}) calculated
neglecting the relativistic corrections.
They are obtained by the Fourier transform of Eq.(\ref{rcd}), where
the charge densities are calculated with the use of
the AV8'+$3N$ poential\cite{He4}.

As pointed out in Ref.\cite{am}, 
the remarkable differences between the black and red curves in Fig. \ref{fig_form} are
the shift of the zero points
and the height of the second peaks in both ${^3}$H and $^{3}$He.
These differences have been understood mainly
by contributions from the MEC,
as mentioned in \S \ref{intro}\cite{am,sch,riska,sickr}.
The relativistic corrections discussed in the previous section
are too small for explaining these differences in the form factors. 

The blue curves, $F_{c,{\rm FB}}$, in Fig. \ref{fig_form} represent
the form factors obtained by the FB charge densities in Ref.\cite{vries}.
On the one hand, because the cut-off parameter, $R$, for $^3$H in the FB-analysis
is small as $3$ fm\cite{vries},
the agreement of the FB-results with the SOG-ones is not good, after the zero-point.
On the other hand, that for $^3$He is taken to be fairly large as $5$ fm\cite{vries}.
It is equal to the maximum value of radii, $R_i$, where the Gaussian is placed
in the SOG-analysis\cite{am}, so that
the FB-form factor is in a good agreement with the SOG-one.

\begin{figure}[ht]
\begin{minipage}[t]{7.2cm}
%
\includegraphics[scale=0.9]{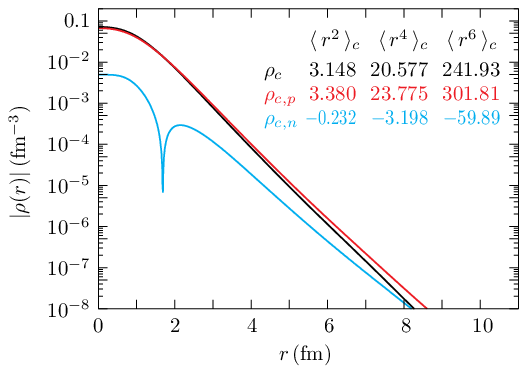}%
\hspace*{-1.5cm}(a)

\end{minipage}
\hspace{0.8cm}
\begin{minipage}[t]{7.2cm}
%
\includegraphics[scale=0.9]{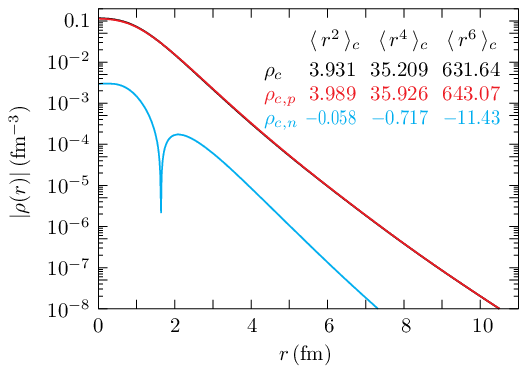}%
\hspace*{-1.5cm}(b)

\end{minipage}
\caption{
The absolute values of the proton(red), neutron(blue), and total charge (black) distributions
in $^{3}$H on the left-hand side  and in $^3$He on the right-hand side.
The distributions are shown in the logarithmic scale
as a function of $r$ in the linear scale.
The values of the corresponding $n$th moment, $\avr{r^n}$, are
also given in the each figure in units of fm$^n$. 
For details, see the text.
}
\label{cpcn}
\end{figure}

Before discussing how the MEC-contributes to the charge distributions
explain the differences between the AV8'+$3N$- and experimental form factors,
let us present the {\it point} nucleon and charge distributions obtained by the AV8'+$3N$-claculations.
In Fig.\ref{point} are shown the point proton(red) and neutron(blue) distributions
as a function of $r$, $r$ being the distance from the center of the nucleus
in $^3$H and $^3$He.
The blue and the red curves show clearly that they are for mirror nuclei.
The characteristic aspects are the distributions around $r\le 1$ fm which are not flat.
These features have been pointed out in $^3$H by Ref.\cite{friar}
and in $^3$He by Ref.\cite{strueve},
using different nuclear interactions.
Ref.\cite{friar} has shown furthermore that the dip of the point proton
distribution in $^{3}$H is smeared out in the charge distribution,
owing to the proton-size\cite{friar}, leaving the top at $r=0$, as in the
black curves in Fig. \ref{point}. Such a smearing to produce a top at $r=0$
is also seen for $^3$He in Fig. \ref{point}.
It will be shown later that these peaks at $r=0$ play an essential role in explaining
the $q$-dependence of the experimental form factors in $^3$H and $^3$He,
cooperating with MEC.
The values of the
$n$th moment, $\avr{r^n}$, are also given in each figure in units of fm$^n$. 
The corresponding moments as mirror nuclei are fairly different from each other.
For example, the msr of the neutron distribution in $^3$H is $3.102$ fm$^{-2}$,
while that of the proton distribution in $^3$He $3.220$ fm$^{-2}$.

The proton(red), neutron(blue) and total(black) charge distributions calculated
with AV8'+$3N$ potential are shown for
$^3$H and $^3$He in Fig. \ref{cpcn}.
They are shown by the absolute values, because
the second bump of the neutron charge distributions is negative,
giving $\int dr^3 \rho_{cn}(r)=0$.
The black curves are the same as those in Fig. \ref{point},
but are shown in the logarithmic scale.
It seems that the neutron contribution to the total charge distributions is negligible.
As a result, the neutron contribution to form factors
has not been discussed so far, as far as the authors know.
The tail of the neutron distribution in $^3$H, however, should be compared to that
of the proton one. 
The values of the $n$th moment of the each charge distribution are
given by the corresponding colors in these figures in units of fm$^n$. 
It is seen that the neutrons yield non-negligible and negative contributions
to the moments of the charge distributions. 
For example, the neutron charge distribution contributes to $\avr{r^6}_c$
by about $25\%$ in $^3$H.  
This is an example to show that the higher moments depend on $\rho_{cn}(r)$,
as mentioned in \S \ref{intro}.

\begin{figure}[ht]
\begin{minipage}[t]{7.2cm}
\includegraphics[scale=0.9]{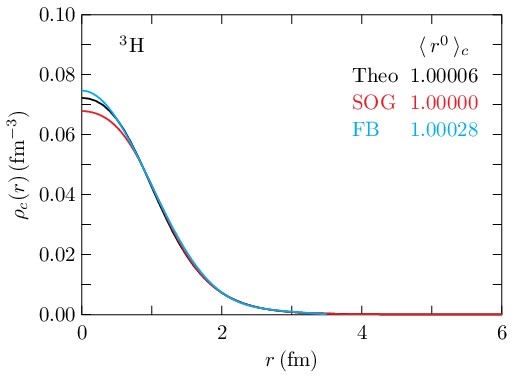}%
\hspace*{-1.5cm}(a)
\end{minipage}\hspace{0.8cm}%
 \begin{minipage}[t]{7.2cm}
  \includegraphics[scale=0.9]{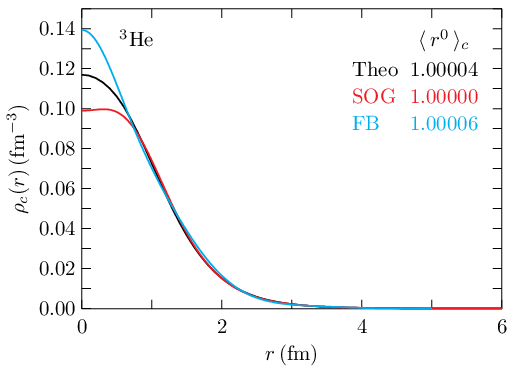}%
\hspace*{-1.5cm}(b)
 \end{minipage}
 
\vspace{3mm} 
 \begin{minipage}[t]{7.2cm}
  \includegraphics[scale=0.9]{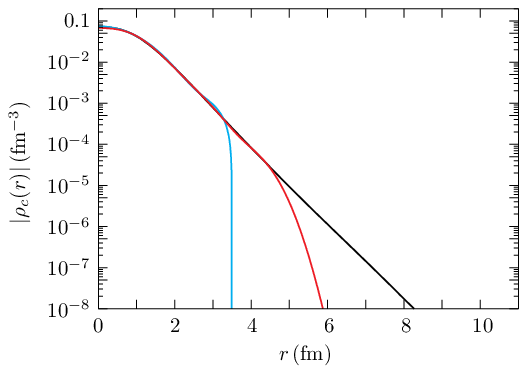}%
\hspace*{-1.5cm}(c)
\end{minipage}\hspace{0.8cm}%
\begin{minipage}[t]{7.2cm}
 \includegraphics[scale=0.9]{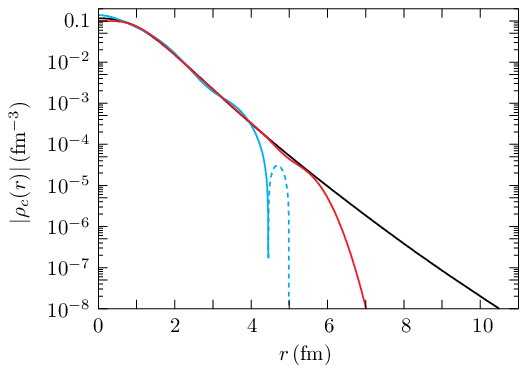}%
\hspace*{-1.5cm}(d)
\end{minipage}
 
\caption{
The charge distributions of $^3$H on the left-hand side
and of $^3$He on the right-hand side, as a function of $r$.
 The red curves are obtained by the sum-of-Gaussians(SOG)-analysis of
the experimental data\cite{am},
the blue ones by the Fourier-Bessel analysis in Ref.\cite{vries},
and the black ones by the present
calculations with AV8'+$3N$ potential.
The figures, (a)and (b), are shown in the linear scale, while (c) and(d)
in the logarithmic scale.
 The dashed curve in (d) shows the negative distribution in the FB-one,
 which is not due to the neutron charge density, but to the oscillation
 of the FB-density\cite{ts}. 
For details, see the text.
}
\label{charge}
\end{figure}

Now, we compare the calculated charge distributions with those
by the SOG- and the FB-analysis.
In Fig. \ref{charge} are shown the charge distributions of $^3$H on the left-hand side
and of $^3$He on the right-hand side.
The red curves are obtained by the SOG-analysis of
the experimental data\cite{am},
the blue ones by the FB-analysis in Ref.\cite{vries},
and the black ones by the present
calculations with AV8'+$3N$ potential in Fig. \ref{cpcn}. As mentioned before,
the red curves by SOG-analysis
reproduce the experimental form factors up to $q^2=30$ fm$^{-2}$,
as in Fig. \ref{fig_form}.
It is seen that in the linear-scale figures, (a) and (b), the clear differences
between three curves appear in the region about $r\le 1$ fm,
while in the logarithmic scale figures, (c) and (d), in the tail above $r=4$ fm. 
Figure \ref{fig_form} has shown that the FB form factor explains well the SOG-one in $^3$He,
but their charge densities are fairly different from each other,
in the both region, $r<1$ fm and $r>4$ fm.
In particular, the FB-density in the region, $r>4$ fm, oscillates with no physical meaning. 
Indeed, this region, $r>4$ fm, are essential for discussions of the
moments, as seen in the next section.
The oscillation of the tails in the FB charge distributions has been pointed out
in $^{48}$Ca and $^{208}$Pb in Ref.\cite{ts}.
Ref.\cite{sickex} discussed some examples, which explain the form factor,
but yield the corresponding charge density with no physical meaning.
The present FB form factors seem to be another example. 
Hence, from now on, we will compare the results of the AV8'+$3N$-calculations mainly
to the SOG-ones. 

Let us explore whether or not
the disagreement between the SOG form factors and the AV8'+$3N$-ones
in Fig. \ref{fig_form} is attributed to the difference between the charge densities
in the region, $r\le 1$ fm and/or $r\ge 4$ fm, in Fig. \ref{charge}.
For this purpose,
first, we define the form factor, $F_{c,{\rm Corr}}(q)$,
by using the AV8'+$3N$ form factor, $F_{c,{\rm Theo}}(q)$, in Fig. \ref{fig_form}.
It will be written so as to reproduces almost the SOG-form factor, $F_{c,{\rm SOG}}(q)$.  
Next, the differences between these two form factors will be studied
in the coordinate space.


We define the difference between $F_{c,{\rm Corr}}$ and $F_{c,{\rm SOG}}$ by
\begin{equation}
\chi^2(q)=\sum_{i=1}^N\frac{1}{\Delta_i^2}
\left(F_{c,{\rm Corr}}(q_i)-F_{c,{\rm SOG}}(q_i)\right)^2 \label{chi}
\end{equation}
where 
\begin{equation}
F_{c,{\rm Corr}}(q_i)=F_{c,{\rm Theo}}(q_i)+\Delta F_c(q_i),
 \quad \Delta_i^2=|F_{c,{\rm SOG}}(q_i)|^2/100.\label{corrff}
\end{equation}
The function, $\Delta F_c(q)$, is expected to stand for
the MEC-contributions to $F_{c,{\rm Theo}}$.
It should satisfy $\Delta F_c(q=0)=0$, to keep the
charge number, $Z$, as
\begin{equation}
\int dr^3 \Delta\rho_c(r)=0 \quad {\rm for}\quad
\Delta\rho_c(r)=\frac{1}{(2\pi)^3}\int dq^3\Delta F_c(q)\exp(-i\vct{q}\cdot\vct{r}).
\label{trans}
\end{equation}
In order to make the structure of $\Delta F_c(q)$ transparent,
we assume that it is written as
\begin{equation}
\Delta F_c(q) = \sum_{m=1}^MC_mq^{2m}\exp(-\lambda^2q^2), \label{deltaff}
\end{equation}
which satisfies Eq.(\ref{trans}). The coefficients, $C_m$, are obtained by
minimizing the value of $\chi^2$,
while the value of $\lambda$ is fixed as $\lambda=0.46188$ fm,
referring to the  SOG-method\cite{am}.
The denominator, $\Delta^2_i$ , in Eq.(\ref{chi}) is chosen, by taking into account
the experimental errors of the electron-scattering cross sections to provide the SOG-values
to be about a few $\%$.
Although the SOG-analysis in Ref.\cite{am} was performed for the values of $q$ between about  
0.2\cite{sza,beck} and 5.5 fm$^{-1}$, we take $q_i=0.5i$ fm$^{-1}$, $i=1,2,\cdots, N=11$.
It will be seen below that these choices of $q$-values are enough not only for discussing
the zero point and the maximum value of the second peak in the form factor,
but also for clarifying later the defect in the tail of the SOG charge distribution.

\begin{figure}[ht]
\includegraphics[scale=0.9]{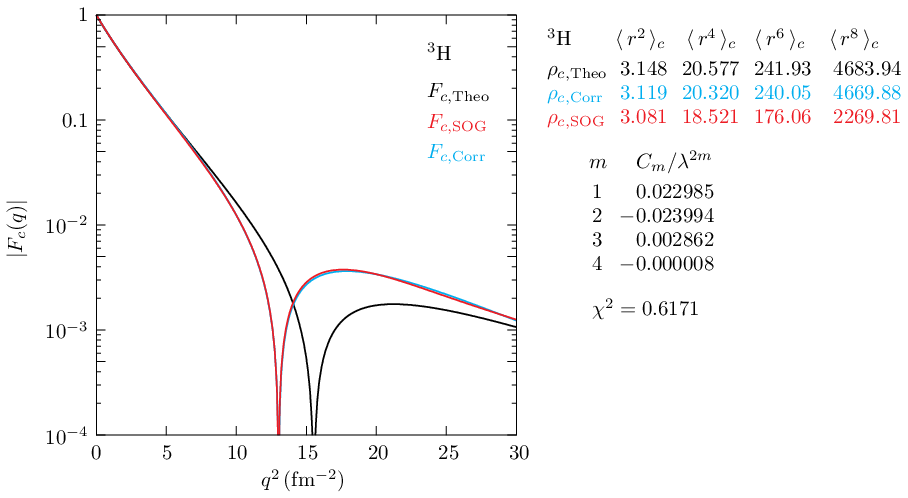}%

\caption{The form factor of $^3$H as a function of the momentum transfer squared, $q^2$.
The form factor, $F_{c,{\rm Theo}}$(black), is calculated with AV8'+$3N$ interaction,
$F_{c,{\rm SOG}}$(red)
obtained by the sum-of-Gaussians(SOG)-analysis of the experimental data\cite{am},
and $F_{c,{\rm Corr}}$(blue) by Eq.(\ref{corrff}). The values of moments
of the charge distributions are listed on the right-hand side by the corresponding colors, and
the values of $C_m$  for Eq.(\ref{deltaff}) are also listed together with the value of $\chi^2$
in Eq.(\ref{chi}).  
For details, see the text.
}
\label{corrH3}
\end{figure}

\begin{figure}[ht]
\includegraphics[scale=0.9]{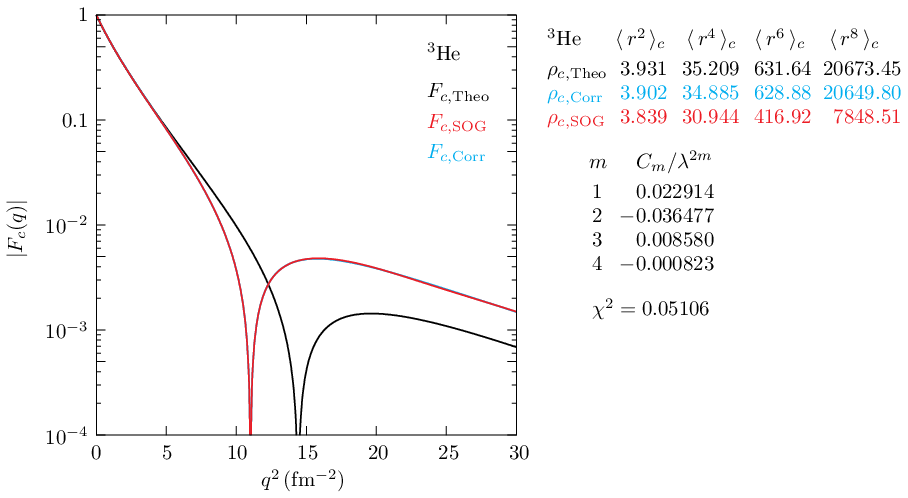}%

\caption{The same as the previous figure, but for $^3$He.
For details, see the text.
}
\label{corrHe3}
\end{figure}

The results are shown in Figs. \ref{corrH3} and \ref{corrHe3}
by the blue curves which are almost on the red curves of $F_{c,{\rm SOG}}$. 
They show that the position of the minimum
and the height of the second peak in $F_{c,{\rm SOG}}$
are well reproduced in $F_{c,{\rm Corr}}$ with $\Delta F_c(q)$.
The values of the coefficient, $C_m$, divided by $\lambda^{2m}$
are listed on the right-hand side in each figure.
The sum of $m$ up to $4$ is enough for simulating the MEC contributions
to the form factors included in the SOG-results\cite{am}.

\begin{figure}[ht]
\begin{minipage}[t]{7.2cm}
\includegraphics[scale=0.9]{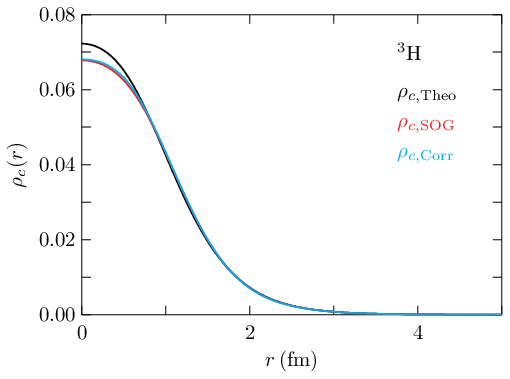}%
\hspace*{-1.5cm}(a)
\includegraphics[scale=0.9]{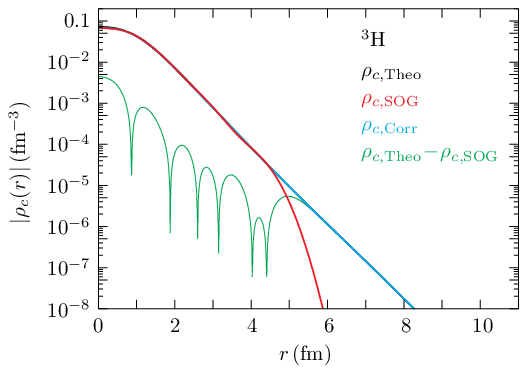}%
\hspace*{-1.5cm}(c)
\end{minipage}\hspace{0.8cm}%
\begin{minipage}[t]{7.2cm}
\includegraphics[scale=0.9]{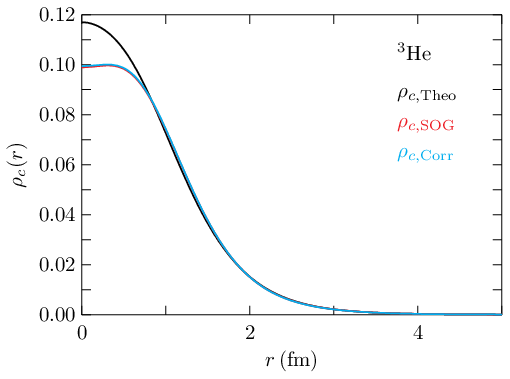}%
\hspace*{-1.5cm}(b)
\includegraphics[scale=0.9]{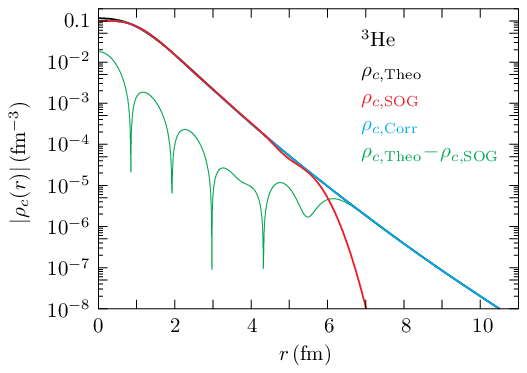}%
\hspace*{-1.5cm}(d)
\end{minipage}
\caption{
 The charge distributions of $^3$H on the left-hand side
and of $^3$He on the right-hand side, as a function of $r$.
 The black curves($\rho_{\rm c,Theo}$) are obtained by the
calculations with AV8'+$3N$ potential, while
the red ones($\rho_{\rm c,SOG}$) by the sum-of-Gaussians(SOG)-analysis of
the experimental data\cite{am}.
The blue ones($\rho_{\rm c,Corr}$) show the charge distributions provided
by the corrected form factors, Eq.(\ref{corrff}).
The upper figures are depicted with the linear scale, while the lower ones
with the logarithmic scale. The green curves are given for reference.
For details, see the text.
}
\label{corrdenc}
\end{figure}

\begin{figure}[ht]
\begin{minipage}[t]{7.2cm}
%
\includegraphics[scale=0.9]{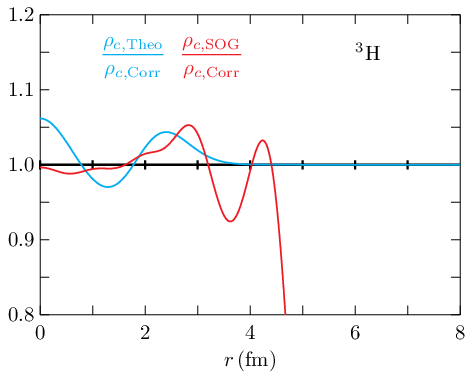}%
\hspace*{-1.5cm}(a)

\end{minipage}
\hspace{0.8cm}
\begin{minipage}[t]{7.2cm}
%
\includegraphics[scale=0.9]{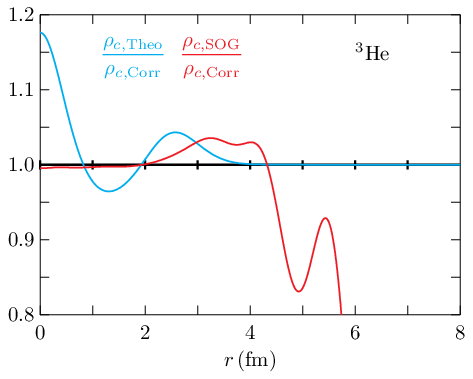}%
\hspace*{-1.5cm}(b)

\end{minipage}
\caption{
 The ratio of the charge distributions, as a function of $r$, in $^3$H(a) and $^3$He(b).
 The blue curve shows
the ratio of the AV8'+$3N$ charge distribution($\rho_{\rm c,Theo}$)
 to the corrected charge distribution($\rho_{\rm c,Corr}$), while the red curve
 that of the SOG charge distribution($\rho_{\rm c,SOG}$) to $\rho_{\rm c,Corr}$.
For details, see the text.
}
\label{rcorrdenc}
\end{figure}

\begin{figure}[ht]
\begin{minipage}[t]{7.2cm}
\includegraphics[scale=0.9]{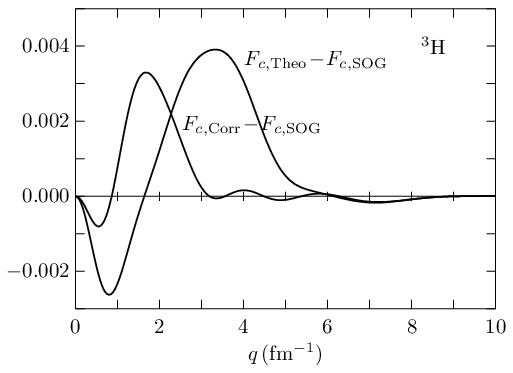}%
\hspace*{-1.5cm}(a)
\end{minipage}\hspace{0.8cm}%
 \begin{minipage}[t]{7.2cm}
  \includegraphics[scale=0.9]{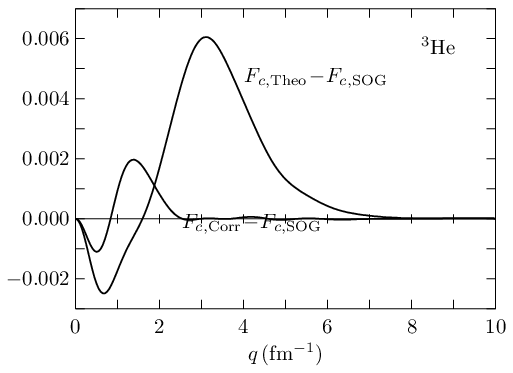}%
\hspace*{-1.5cm}(b)
 \end{minipage}
 
\vspace{3mm} 
 \begin{minipage}[t]{7.2cm}
  \includegraphics[scale=0.9]{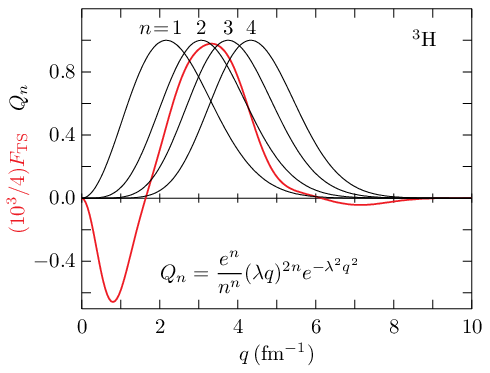}%
\hspace*{-1.5cm}(c)
\end{minipage}\hspace{0.8cm}%
\begin{minipage}[t]{7.2cm}
 \includegraphics[scale=0.9]{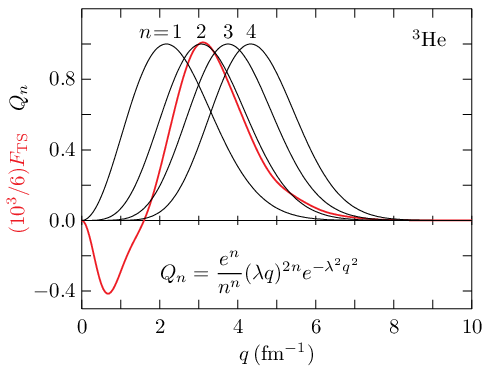}%
\hspace*{-1.5cm}(d)
\end{minipage}
 
\caption{The differences between various form factors, as a function of $q^2$, in $^3$H(a)
 and $^3$He(b), and the components of $\Delta F_c(q)$ in $^3$H(c) and $^3$He(d).
The red curves in (c) and (d) show $F_{\rm TS}=F_{c,{\rm Theo}}-F_{c,{\rm SOG}}$ in (a) and (b),
 respectively.
For details, see the text.
}
\label{gauss}
\end{figure}

Next, we transform $F_{c,{\rm Corr}}(q)$
into the charge distribution($\rho_{c,{\rm Corr}}(r)$), as in Eq.(\ref{trans}).
In Fig. \ref{corrdenc}, $\rho_{c,{\rm Corr}}(r)$
is depicted by the blue curve with the linear and logarithmic scales.
The black($\rho_{c,{\rm Theo}}(r)$) and red($\rho_{c,{\rm SOG}}(r)$)
curves show the charge densities
obtained by the AV8'+$3N$-calculation and the SOG analysis\cite{am},
respectively, as in Fig. \ref{charge}.
It is clearly seen that $\Delta F_c(q)$ works to depress the AV8'+$3N$-distribution
around $r<1$ fm 
in upper figures and yields no effect on the tail in the lower figures
in both $^3$H and $^3$He.
In the $q$-space, the difference between $F_{c,{\rm SOG}}$
and $F_{c,{\rm Theo}}$ has been explained
by the negative contribution of the mesonic degrees of freedom\cite{am,sch,riska,sickr}.
In the $r$-space,
the MEC plays a role to suppress the bump near the center
of the charge density calculated by the nucleonic degrees of freedom.
These facts are displayed in a different way in Fig. \ref{rcorrdenc},
which shows the ratios, $\rho_{c,{\rm Theor}}(r)/\rho_{c,{\rm Corr}}(r)$(blue)
and $\rho_{c,{\rm SOG}}(r)/\rho_{c,{\rm Corr}}(r)$(red)
as a function of $r$ in $^3$H(a) and $^3$He(b). 
On the one hand, the difference between $\rho_{c,{\rm Theor}}(r)$ and $\rho_{c,{\rm Corr}}(r)$
seen up to $r\approx 4$ fm is due to $\Delta \rho_c(r)$ coming from $\Delta F_c(q)$,
and after $r\approx 4$ fm, these two distributions are almost equal to each other.
The bump in the region $r\le 1$ fm in $\rho_{c,{\rm Theor}}(r)$
is smeared out by $\Delta \rho_c(r)$.
On the other hand, the red curves show that
$\rho_{c,{\rm SOG}}(r)$ is almost equal to $\rho_{c,{\rm Corr}}(r)$
up to $r\approx 2$ fm, but deviates from it after $r\approx 2$ fm, and approaches to zero,
implying that the tails of $\rho_{c,{\rm SOG}}(r)$ is not described as those of
$\rho_{c,{\rm Corr}}(r)$ and $\rho_{c,{\rm Theo}}(r)$ calculated in a quantum mechanical way.

The above effects of $\Delta F_c(q)$ on the charge distribution near the center
and tail are understood analytically on the basis of Eq.(\ref{deltaff}).
In Fig. \ref{gauss}-(a) and -(b) are shown for $^3$H and $^3$He, respectively,
the differences between $F_{c,{\rm Theo}}(q)$ and $F_{c,{\rm SOG}}(q)$
and between $F_{c,{\rm Corr}}(q)$ and $F_{c,{\rm SOG}}(q)$. 
In Figs. \ref{gauss}-(c) and  \ref{gauss}-(d),
the components of $\Delta F_c(q)$ are shown in terms of $Q_n(q)$, whose maximum value
at $q^2=n/\lambda^2$ is normalized to be $1$ as
\begin{equation}
Q_n(q)=\frac{e^n}{n^n}(\lambda q)^{2n}\exp (-\lambda^2q^2).
\end{equation}
For reference, the $q$-dependences of $F_{\rm TS}(q)=F_{c,{\rm Theo}}(q)-F_{c,{\rm SOG}}(q)$
are also shown by the red curves in Figs. \ref{gauss}-(c) and  \ref{gauss}-(d).
The correction, $\Delta F_c(q)$, is given by $\Delta F_c(q)=F_{\rm CS}(q)-F_{\rm TS}(q)$,
where $F_{\rm CS}(q)=F_{c,{\rm Corr}}(q)-F_{c,{\rm SOG}}(q)$ in the upper figures.
Comparing Fig. \ref{gauss}-(a) to -(c), or -(b)) to (d), and the weights of the components,
$C_m/\lambda^{m}$, listed in Figs. \ref{corrH3} and \ref{corrHe3},
it is seen that the main components of $\Delta F_c(q)$ are given by $Q_1(q)$ and $Q_2(q)$.
The Fourier transforms of the corresponding components, $m=1$ and $2$ in Eq.(\ref{deltaff}),
provide the main corrections to the charge density, respectively, as
\begin{eqnarray}
\Delta \rho_{c,1}(r)&=&\frac{Z}{32\pi^{3/2}\lambda^3}\left(\frac{C_1}{\lambda^2}\right)
 \left[6-\left(\frac{r}{\lambda}\right)^2\right]\exp (-\frac{r^2}{4\lambda^2}),\label{g1}\\
\Delta \rho_{c,2}(r)&=&\frac{Z}{128\pi^{3/2}\lambda^3}\left(\frac{C_2}{\lambda^4}\right)
 \left[60-20\left(\frac{r}{\lambda}\right)^2
  +\left(\frac{r}{\lambda}\right)^4\right]\exp (-\frac{r^2}{4\lambda^2}). \label{g2}
 \end{eqnarray}
The sum of $\Delta \rho_{c,1}(r)$ and $\Delta \rho_{c,2}(r)$ yields a negative contribution
depressing the center of the charge distribution, and almost no effects
on the tails in both $^3$H and $^3$He.
These understandings of $\Delta F_c(q)$ are consistent with those reported
in the previous papers\cite{kat,mc,mc2}. 
On the one hand, in 1970, Refs.\cite{mc, mc2} stressed the necessity of the additional charge
distribution near the center in order to reproduce the diffraction minimum
in the experimental position in the $^3$He form factor, although the main distribution
is assumed to have the tail of the Gauss function.
On the other hand, in 1982,
Ref.\cite{kat} showed that effects of the MEC
appear around the center($r<1$ fm) of the charge distribution in $^4$He. 

We note that the slight difference between the moments of $\rho_{c,{\rm Theo}}$
and $\rho_{c,{\rm Corre}}$ listed in Fig. \ref{corrHe3} is explained
by the above two equations. 
For example, Fig. \ref{corrHe3} lists $3.931(\rho_{c,{\rm Theo}})$
and $3.902(\rho_{c,{\rm Corre}})$ fm$^2$ for $\avr{r^2}_c$.
Eq.(\ref{g1}) for $^3$He provides the value of the msr to be $-0.029$,
which is equal to
($3.902(\rho_{c,{\rm Corr}})-3.931(\rho_{c,{\rm Theor}}))=-0.029$.
The value of the msr of the distribution given by Eq.(\ref{g2}) is exactly $0$.
The values of $\avr{r^4}_c$ given by Eqs.(\ref{g1})
and (\ref{g2}) are $-0.125$ and $-0.199$, respectively,
whose sum is $(-0.125-0.199) = -0.324$,
while the values in Fig. \ref{corrHe3} is the same
as $(34.885(\rho_{c,{\rm Corr}})-35.209(\rho_{c,{\rm Theo}}))=-0.324$.

\begin{figure}[ht]
\includegraphics[scale=0.9]{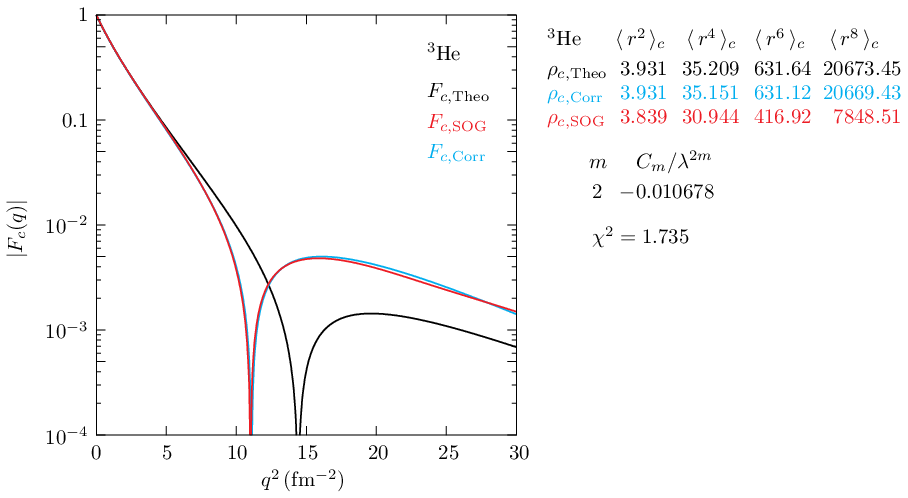}%

\caption{The form factor of $^3$He as a function of the momentum transfer squared, $q^2$.
The form factor, $F_{c,{\rm Theo}}$(black), is calculated with AV8'+$3N$ interaction,
$F_{c,{\rm SOG}}$(red)
obtained by the sum-of-Gaussians(SOG)-analysis of the experimental data\cite{am},
and $F_{c,{\rm Corr}}$(blue) by Eq.(\ref{corrff}) with the $\Delta F_c(q)$ in
Eq.(\ref{deltaff}) with $m=2$-component only,
as indicated on the right-hand side of the figure.
 The values of moments,$\avr{r^n}_c$,
of the charge distributions are listed
by the corresponding colors
in units of fm$^n$, and the values of $C_{m=2}$ for Eq.(\ref{deltaff}) is also
listed together with the value of $\chi^2$ in Eq.(\ref{chi}).  
For details, see the text.
}
\label{22corrHe3}
\end{figure}

\begin{figure}[ht]
\begin{minipage}[t]{7.2cm}
%
\includegraphics[scale=0.9]{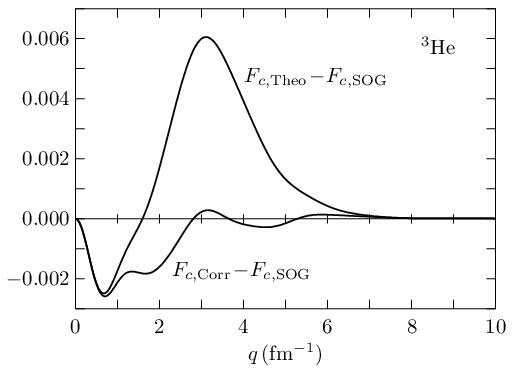}%
\hspace*{-1.5cm}(a)

\end{minipage}
\hspace{0.8cm}
\begin{minipage}[t]{7.2cm}
%
\includegraphics[scale=0.9]{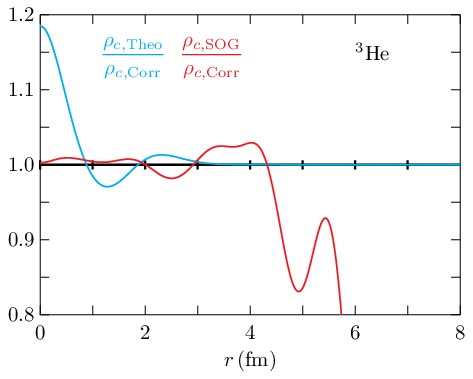}%
\hspace*{-1.5cm}(b)

\end{minipage}
\caption{
The differences between form factors in $^3$He and the ratios
 between the charge distributions.
In (a), the form factors, $F_{c,\rm Theor}, F_{c,\rm Corr}$ and $F_{c,\rm SOG}$
denote the AV8'+$3N$-,  AV8'+$3N$(corrected by the $m=2$-component)-, and the SOG-one, respectively,
while in (b), the corresponding charge distributions are written as
$\rho_{c,\rm Theor}, \rho_{c,\rm Corr}$ and $\rho_{c,\rm SOG}$, respectively. 
For details, see the text.
}
\label{22rcorrdencHe3}
\end{figure}

Actually, Eq.(\ref{g2}) only is enough to reproduce, for example, Fig. \ref{corrHe3},
as shown in Fig. \ref{22corrHe3} by changing the value of $C_2$.
The blue curve exhibits almost the same minimum-position
and the same maximum-value of the second peak
in the corrected form factor, $F_{c,{\rm Corr}}$,
as in the SOG-one(red curve).
As mentioned before, the $m=2$-component of $\Delta F_c(q)$ becomes minimum
around $q=\sqrt{2}/\lambda \approx 3.06$ fm$^{-1}$, where the MEC-contribution
becomes important\cite{sch,kat,str,riska,sickr}.
In this case, there is no change in the msr-value, as mentioned before,
and almost no change in the values of the higher moments,
between $\rho_{\rm c,Theor}$ and  $\rho_{\rm c,Corr}$, as listed in Fig. \ref{22corrHe3}.


Up to this stage, the disagreement of the AV8'+$3N$ form factors with the SOG ones, shown
in Fig. \ref{fig_form}, is attributed to the difference between the heights
of their charge distributions around $r<1$ fm,
but not to the difference between the tails.
The difference between the tails, however, should affect 
the form factors, $F_{c,{\rm Corr}}$ and $F_{c,{\rm SOG}}$, in other ways. 
In Fig. \ref{gauss}-(a) and -(b) are shown for $^3$H and $^3$He, respectively,
the differences between $F_{c,{\rm Theo}}$ and $F_{c,{\rm SOG}}$ and between $F_{c,{\rm Corr}}$
and $F_{c,{\rm SOG}}$ in Figs. \ref{corrH3} and \ref{corrHe3}. 
They show that the main peak of $F_{\rm TS}=F_{c,{\rm Theo}}-F_{c,{\rm SOG}}$ around $q >3$ fm$^{-1}$ 
disappears in $F_{\rm CS}=F_{c,{\rm Corr}}-F_{c,{\rm SOG}}$ by $\Delta F_c(q)$,
but that $F_{\rm CS}\ne0$ in the region, $q<3$ fm$^{-1}$.
The peak near $q=2$ fm$^{-1}$ in $F_{\rm CS}$ is mainly due to the $m=1$-component
of $\Delta F_c(q)$, because it disappears 
as shown in Fig. \ref{22rcorrdencHe3}-(a) which is calculated by the $m=2$ component only,
as in Fig. \ref{22corrHe3}.
Fig. \ref{22rcorrdencHe3}-(a) shows that the $m=2$-component 
only is enough for reproducing the SOG form factor above $q>3$ fm$^{-1}$,
leaving the first dip almost in the same way as in $F_{\rm TS}$.
This fact implies that the difference appearing in the region, $q<1$ fm$^{-1}$,
in $F_{\rm CS}$ is owing to the difference of the tails
between $\rho_{c,{\rm Corr}}(r)$ and $\rho_{c,{\rm SOG}}(r)$,
as shown more concretely below.

According to Eq.(\ref{npfm}), the form factor, $F_c(q)$, is expanded in terms of the moments as
\begin{equation}
F_c(q)=\sum_{m=0}^\infty(-1)^m\frac{1}{(2m+1)!}q^{2m}\avr{r^{2m}}_c. \label{pol}
\end{equation}
There is no constraint on the expansion in a physical point of view,
but in a practical use with
the limited number of terms,
we should know its convergence and the magnitude of the reminder term.   
For this purpose, the Leibnitz criterion is useful, which requires the sufficient
condition for the convergence of the alternating series,  
\begin{equation}
\frac{1}{(2m+1)!}q^{2m}\avr{r^{2m}}_c>\frac{1}{(2m+3)!}q^{2(m+1)}\avr{r^{2(m+1)}}_c\,
\,\,\,\,\,\,(\rightarrow 0 \,\,\,{\rm for} \,\,\,m \rightarrow \infty).\label{leib}
\end{equation}
In this case, for the sum of the limited number of the terms:
\begin{equation}
F_k(q)=\sum_{m=0}^k(-1)^m\frac{1}{(2m+1)!}q^{2m}\avr{r^{2m}}_c\,, \label{ppol}
\end{equation} 
there is the alternating-series-estimation-theorem as 
\begin{equation}
|F_k(q)-F_c(q)| \le \frac{1}{(2m+3)!}q^{2(m+1)}\avr{r^{2(m+1)}}_c,\label{les}
\end{equation}
which gives an approximate magnitude on the sum of the neglected terms.

In using the above expansion, for example, at $q=0.5$ fm$^{-1}$,
the values of the moments listed in Fig. \ref{corrHe3}
for $^3$He provide the first $5$ terms up to the eighth moment of Eq.(\ref{ppol}) as
\begin{equation}
F_{c,{\rm Corr}}(q=0.5)\approx 1-0.16258+0.01817-0.00195+0.00022=0.85386,\label{fv1}
 \end{equation}
and
 \begin{equation}
F_{c,{\rm SOG}}(q=0.5)\approx 1-0.15996+0.01612-0.00129+0.00008=0.85495.\label{fv2}
 \end{equation}
The above two equations give $F_{\rm CS}(q=0.5)=-0.0011$, which is almost the same
as the one in Fig. \ref{gauss}-(b).
Indeed, this value is obtained by the numerical calculation of $F_{\rm CS}$
without the above expansions.
In the same way, we have $F_{\rm TS}(q=0.5)=-0.0021$ for the values
listed in Fig. \ref{corrHe3},
corresponding to the value in Fig. \ref{gauss}-(b).
The sixth terms neglected in Eqs.(\ref{fv1}) and (\ref{fv2}) are small enough
for the present arguments, according to Eq.(\ref{les}).
Thus, the dips of $F_{\rm CS}(q)$ and $F_{\rm TS}(q)$ in the small $q$ region
in Fig. \ref{gauss}-(b) are understood
in terms of the moments of $\rho_{\rm c,Corr}(r)$, $\rho_{\rm c,Theo}(r)$
and $\rho_{\rm c,SOG}$
whose values are different from one another.
The first dips of $F_{\rm CS}$ and $F_{\rm TS}(q)$
in Fig. \ref{gauss}-(a) and Fig. \ref{22rcorrdencHe3}-(a) 
are also explained in the same way.
The moments are insensitive to the charge distribution near its center,
but dominated by its tail, because the integrand of the $n$th moment is given by
$r^{n+2}\rho_c(r)$.
This fact implies that the difference appearing in the region, $q<1$ fm$^{-1}$
in $F_{\rm CS}$ is owing to the difference of the tails
between $\rho_{c,{\rm Corr}}(r)$ and $\rho_{c,{\rm SOG}}(r)$.

We conclude in this section that
the SOG form factors, $F_{c,{\rm SOG}}(q)$,
in the region of $0.2<q<5.5$ fm$^{-1}$ for $^3$H and for $^3$He
in Ref.\cite{am} are reproduced 
by the corrected AV8'+$3N$-ones, $F_{c,{\rm Corr}}(q)$.
In $F_{\rm c,SOG}$, the mesonic degrees of freedom play essential roles
to reproduce the experimental form factors\cite{am},
while their roles are  simulated by
$\Delta F_c(q)$ added to the AV8'+$3N$-one, $F_{c,{\rm Theo}}(q)$,
 as $F_{c,{\rm Corr}}(q)=F_{c,{\rm Theo}}(q)+\Delta F_c(q)$.
This fact, however, does not imply that the SOG charge distributions
and the corrected AV8'+$3N$-ones are the same as each other above $r\approx 2$ fm,
as shown in Fig.  \ref{rcorrdenc}.
In particular, Fig. \ref{corrdenc} indicates that
the SOG-distributions
do not reproduce
well the tails of the charge distributions above $5$ fm,
which are described in the quantum-mechanical calculations.
The MEC-contributions, expressed by $\Delta F_c(q)$, are not important
for estimation of the moments of the charge distributions,
whereas the tails are expected to dominate the values of the moments.
The difference between the tails is observed through the PWBA form factors
at low $q$ which are expressed in terms of the moments.

\section{Moments of the charge distribution in $^3$H and $^3$He}\label{moment}

\begin{figure}[ht]
\begin{minipage}[t]{7.2cm}
\includegraphics[scale=0.9]{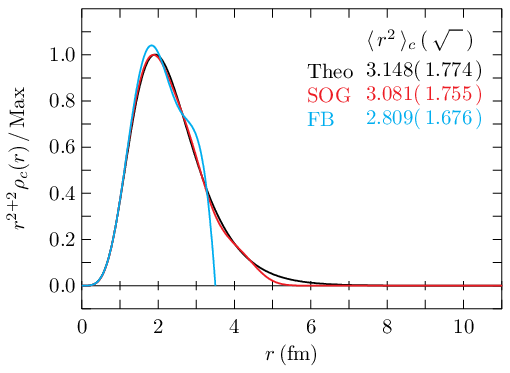}%
\hspace*{-1.5cm}(a)
\end{minipage}\hspace{0.8cm}%
 \begin{minipage}[t]{7.2cm}
\includegraphics[scale=0.9]{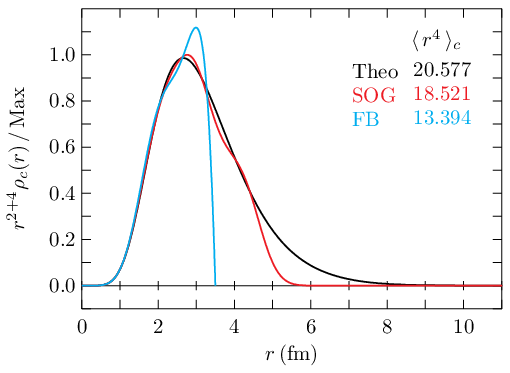}%
\hspace*{-1.5cm}(b)
 \end{minipage}
 
\vspace{3mm} 
 \begin{minipage}[t]{7.2cm}
\includegraphics[scale=0.9]{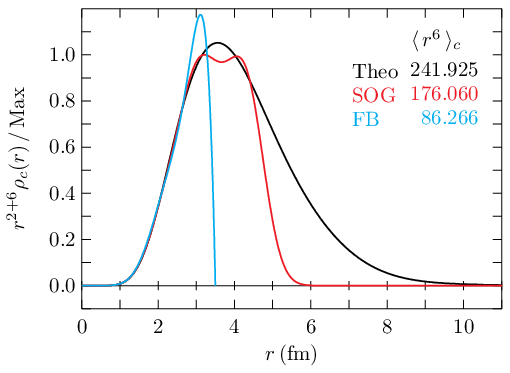}%
\hspace*{-1.5cm}(c)
\end{minipage}\hspace{0.8cm}%
\begin{minipage}[t]{7.2cm}
\includegraphics[scale=0.9]{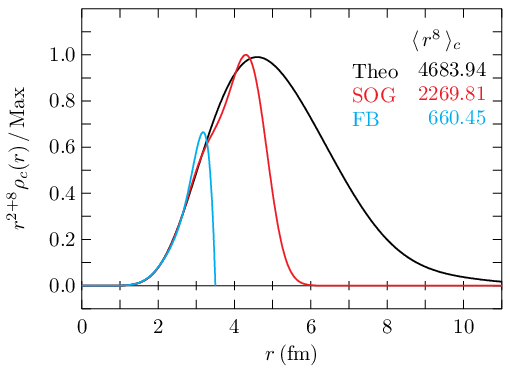}%
\hspace*{-1.5cm}(d)
\end{minipage}
 
\caption{
The integrand, $r^{2+n}\rho_c(r) $, for the $n(\le 8)$th moments of
 the charge distribution in $^3$H,
 as a function of $r$.
The black curves are for the AV8'+$3N$-calculations, the red ones for the SOG-distribution,
and the blue ones for the FB-distribution.
The integrands are divided by Max which indicates the maximum value
 of $r^{2+n}\rho_{c,{\rm SOG}}(r)$.
The value of the each moment, $\avr{r^n}_c$, is given in units of fm$^n$
in the corresponding figure.
For details, see the text.
}
\label{h_moment}
\end{figure}

In this section, we investigate the moments of the charge distribution,
in particular, focusing on the role of the tail.
We compare the moments given by the SOG- and FB-analyses
to those obtained by AV8'+$3N$, but not to the corrected ones,
because the AV8'+$3N$-moments and the corrected ones   
are almost the same as each other, as listed in Figs. \ref{corrH3}, \ref{corrHe3}
and \ref{22corrHe3}.
The comparison of the AV8'+$3N$-moments to the SOG- and FB-ones has another meaning.
Unlike mean-field models\cite{sly4,nl3} for medium and heavy nuclei, 
the AV8'+$3N$-calculations for the three-body systems
do not use the experimental values of the msr, $\avr{r^2}_c$,
as an input for fixing the nuclear interactions\cite{He4}.

Figure \ref{h_moment} shows the integrands as a function of $r$ for calculating
the $n(n\le 8)$th moment of the charge distribution in $^3$H.
The black curves are obtained by the AV8'+$3N$-distribution, while the red(blue)ones with the use
of the SOG(FB)-one. Note that the AV8'+$3N$-results have no relativistic corrections
which are not essential for the present discussions.
The integrands are shown dividing them by the maximum value(Max) of the SOG-one.  

It is seen that the three curves are fairly different from each other.
Even the `experimental values' of the msr obtained by the SOG- and FB- analyses are
different from each other, as listed in the corner of Fig. \ref{h_moment}-(a).
The former are larger than the latter, and both are smaller than the AV8'+$3N$-values. 
This fact is easily understood by Fig. \ref{h_moment} as a result of the different
shapes of the integrands.

On the one hand, in the FB-integrands, the cut-off value of $R=3.5$ fm\cite{vries} dominates
them even for the msr, $\avr{r^2}_c$.
As a result, all the values of $\avr{r^n}_c$ are smaller than those by
the SOG-analyses and the AV8'+$3N$-calculations, as listed in the right-hand side of each figure.

On the other hand, the SOG-values of moments are dominated by the value
of $R_{\rm max}=5$ fm\cite{am}.
The value of the msr is $3.081$ fm$^2$, while that of the AV8'+$3N$-one $3.148$ fm$^2$ as listed
in Fig. \ref{h_moment}-(a) and $3.181$ fm$^2$,
when adding the relativistic correction, $0.033$ fm$^2$,
in Eq.(\ref{msr}).
Ref.\cite{am} provided 3.087(0.302) fm$^2$ from the slope of the SOG-form factor
at $q\rightarrow 0$.
Because of the about $10\%$ error, this value does not contradict 
all other values, including the FB-value, $2.809$ fm$^2$.

The differences between the AV8'+$3N$-, SOG-, and FB-estimations of the moments increase,
with the increasing value of $n$,
reflecting the differences between the tails of the corresponding charge distributions
in Fig. \ref{charge}.
In the case of $\avr{r^4}_c$ which is required for discussions on Zemach
moments\cite{sickz,din}, the AV8'+$3N$-calculation provides $20.578$ fm$^4$ which is
larger about by $1.54$ times than that of the FB-value, $13.394$ fm$^4$,
while the SOG-analysis gives $18.521$ fm$^4$.
In the eighth moment, as in Fig. \ref{h_moment}-(d),
the SOG-value is less than a half of that by the AV8'+$3N$-calculation, and the FB-value less than
one seventh.  


\begin{figure}[ht]
\begin{minipage}[t]{7.2cm}
\includegraphics[scale=0.9]{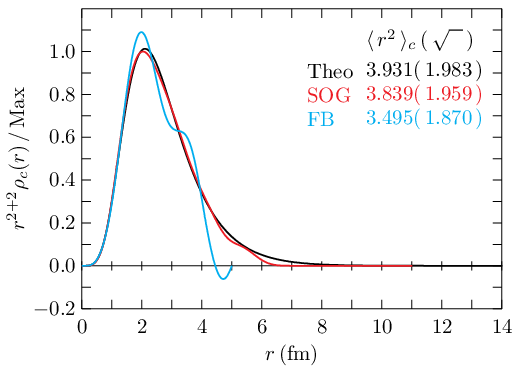}%
\hspace*{-1.5cm}(a)
\end{minipage}\hspace{0.8cm}%
 \begin{minipage}[t]{7.2cm}
\includegraphics[scale=0.9]{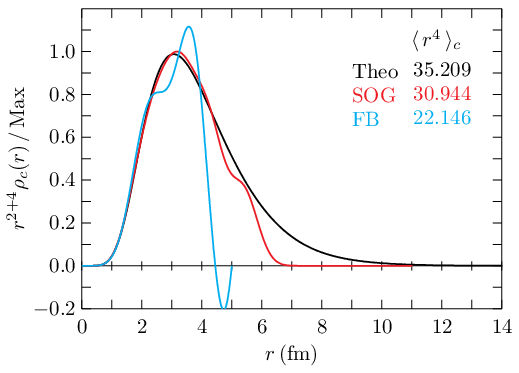}%
\hspace*{-1.5cm}(b)
 \end{minipage}
 
\vspace{3mm} 
 \begin{minipage}[t]{7.2cm}
\includegraphics[scale=0.9]{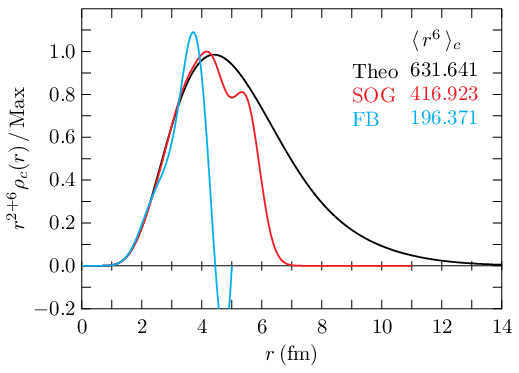}%
\hspace*{-1.5cm}(c)
\end{minipage}\hspace{0.8cm}%
\begin{minipage}[t]{7.2cm}
\includegraphics[scale=0.9]{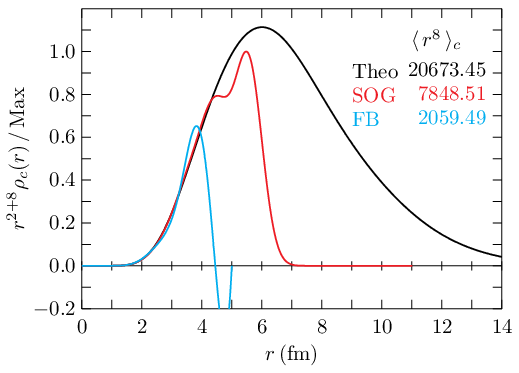}%
\hspace*{-1.5cm}(d)
\end{minipage}
 
\caption{
The integrands for the moments of the charge distribution in $^3$He,
 as a function of $r$.
The black curves are for the AV8'+$3N$-calculations, the red ones for the SOG-distribution,
and the blue ones for the FB-distribution.
The integrands are divided by Max which is the maximum value
 of $r^{2+2n}\rho_{c,{\rm SOG}}(r)$.
 The value of the each moment, $\avr{r^n}_c$ in units of fm$^n$ is given 
in the corresponding figure.
For details, see the text.
}
\label{he_moment}
\end{figure}

\begin{figure}[ht]
 \hspace{-1cm}
\includegraphics[scale=0.9]{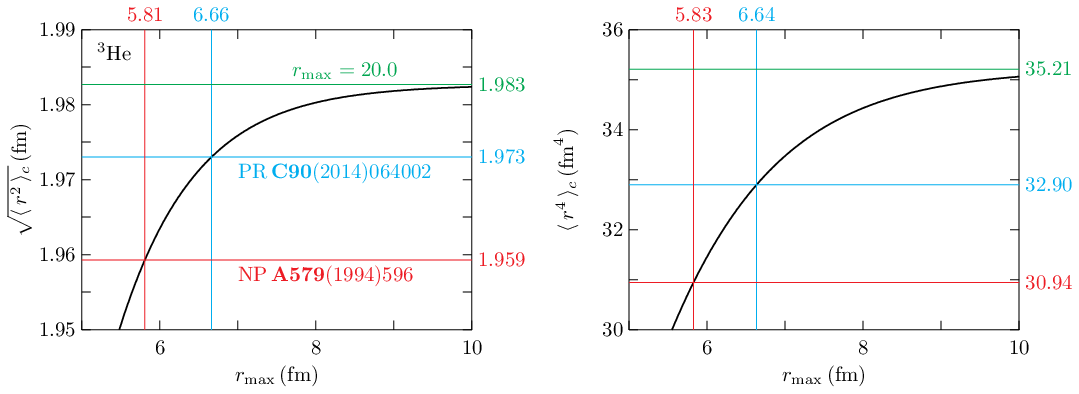}%
\caption{
 The dependence of the values of the moments in the AV8'+$3N$-calculations
 on the cut-off parameter, $r_{\rm max}$, in $^3$He.
The notation, $r_{\rm max}$, denotes the upper limit of the definite integral
in calculating $\avr{r^n}_c$. 
 The black curves show the $r_{\rm max}$-dependences 
 of $\sqrt{\avr{r^2}_c}$ on the left-hand side and of $\avr{r^4}_c$
 on the right-hand side, respectively.
 The red and blue lines indicate the corresponding values of Refs.\cite{am} and \cite{sickz},
 respectively.
 For details, see the text.
}
\label{cutoff}
\end{figure}

Figure \ref{he_moment} shows the integrands to calculate the $n(n\le 8)$th moment
of the charge distribution in $^3$He, in the same way as Fig. \ref{h_moment}. 
It is seen that inadequate descriptions of the tail in the FB- and SOG-analyses induce
the discrepancy between the `experimental' values and the AV8'+$3N$-one, as in $^3$H. 
Ref.\cite{am} has estimated the value of the root msr from the slope
of the form factor at $q\rightarrow 0$,
yielding $1.959(0.030)$ fm, which agrees with the one of the SOG
calculated in the Fig. \ref{he_moment}-(a). The AV8'+$3N$-value, $1.983$ fm,
is larger than that, but within the error, $0.030$ fm.
With respect to $\avr{r^4}_c$, 
Fig.\ref{he_moment}-(b) provides
$30.944$ fm$^4$ for the SOG, while $35.210$ fm$^4$ for the AV8'+$3N$-value.
The new analysis of the SOG in Ref.\cite{sickz} up to $q_{\rm max}=10$ fm$^{-1}$,
taking account of the additional tail correction,
provides the value of $32.90(1.6)$ fm$^4$, together with the value of
$1.973(0.014)$ fm for the root msr.
Both values of $\avr{r^2}_c$ and $\avr{r^4}_c$ approach those of the AV18.
The tail-correction was made by considering the $p$- and $n$-separation energies,
but the charge distribution of the new SOG analysis can not be obtained from Ref.\cite{sickz}.
Instead, we note that
the old and new values of the SOG analyses are reproduced
in the AV8'+$3N$-calculations by introducing almost the same cut-off parameters, $r_{\rm max}$,
for the integration of $\avr{r^2}_c$ and $\avr{r^4}_c$, as in Fig. \ref{cutoff}.
It is seen that the values of $r_{\rm max}$ to obtain the SOG-ones,
as indicated in Fig. \ref{cutoff}, are
almost equal to those expected in Figs. \ref{he_moment}-(a) and (b).
All the analysis-dependences of the moments are thus understood to come from
insufficient descriptions of the tail of the charge distributions,
compared with the AV8'+$3N$-one on the basis of the quantum mechanical calculations.


\section{Discussions}  \label{dis}

\begin{table}
 \hspace*{1cm}%
\begin{tabular}{|l|c|c|c|c|} \hline
$^3$H&
$\avr{r^2}_c$&
$\avr{r^4}_c$&
$\avr{r^6}_c$&
$\avr{r^8}_c$
\\ \hline
\rule{0pt}{12pt}%
AV  & $ 3.181[3.148]$ & $20.925[20.577]$ &$ 246.864[241.925]$ & $4780.271[4683.938]$  \\
SOG  & $ 3.081(0.302)$ & $18.521$ &$ 176.060$ & $2269.81$  \\ 
FB  &  $2.809(0.101)$  &$13.394$  &$86.266$  &$660.45$   \\ 
PA & $ 2.876(0.022)$ & $ 12.612(0.06)$&$58.867(0.151)$ & $185.214(0.217)$  \\   \hline
\end{tabular}
\caption{
The values of the $n$th moments of the charge distributions in $^3$H.
The AV-values are obtained by the present calculations with the AV8'+$3N$-force,
 adding the relativistic corrections to the ones in the square brackets.
Others are by the analyses of the experimental form factors
by the sum-of-Gaussians(SOG) for $0.0477\cite{beck}\le q^2 \le 
30\cite{am}$ fm$^{-2}$,
 the Fourier-Bessel method for $0.260\le q^2\le 8.01$
 fm$^{-2}$\cite{vries},
and the polynomial ansatz(PA) for $0.0477\le q^2\le 2.96$ fm$^{-2}$\cite{beck}.
 The experimental errors are given in the parentheses, if available in the references.
All the values are given in units of fm$^n$.  For the details, see the text. 
}
 \label{h_value}
\end{table}

\begin{table}
 \hspace*{1cm}%
\begin{tabular}{|l|c|c|c|c|} \hline
$^3$He&
$\avr{r^2}_c$&
$\avr{r^4}_c$&
$\avr{r^6}_c$&
$\avr{r^8}_c$
\\ \hline
\rule{0pt}{12pt}%
AV  & $ 3.964[3.931]$ & $35.644[35.209]$ &$ 640.116[631.641]$ & $20925.242[20673.447]$  \\
SOG  & $ 3.839(0.118)$ & $30.944$ &$ 416.923$ & $7848.51$  \\ 
FB  &  $3.495(0.071)$  &$22.146$  &$196.371$  &$2059.49$   \\
PA & $ 3.894(0.120)$ & $ 32.100(4.714)$&$ 404.495(137.295)$ & $5312.563(2887.291)$  \\ 
   \hline
\end{tabular}
\caption{
The values of the $n$th moments of the charge distributions in $^3$He.
The AV-values are obtained by the present calculations with the AV8'+$3N$-force,
 adding the relativistic corrections to the ones in the square brackets.
Others are by the analyses of the experimental form factors
by sum-of-Gaussians(SOG) for $0.0316\cite{sza}\le q^2\le 30$ fm$^{-2}$\cite{am},
the Fourier-Bessel method for $0.0324\le q^2\le 102.1$ fm$^{-2}$\cite{vries},
and the polynomial ansatz(PA) for $0.200\le q^2 \le 3.700$ fm$^{-2}$\cite{ott}.
 The experimental errors are indicated in the parentheses, if available in the
 references.
 All the values are given in units of fm$^n$.
 For the details, see the text. 
}
 \label{he_value}
\end{table}

It seems that there is no way to determine the values of the moments
with high accuracy by improving the SOG- and the FB-method,
even if the world data available at present are employed.
One reason of this fact is that experiments performed so far do not provide
accurate data
at low $q$-region as $q^2 < 0.1$ fm$^{-2}$ which are sensitive to the tail of $\rho_c(r)$.
Another reason is on the SOG- and the FB-method themselves.
They are suitable for investigation of
the gross profile of $\rho_c(r)$, but seem not to be appropriate
for studying the detail of the tail\cite{sog,fl,fb}. 
The SOG-method required additional tail correction
to the Gaussian-damping in Refs.\cite{sickz,sickhe4}, while the FB-one
induced an oscillation of the tail dominated by the cut-off parameter, $R$.  
In this section, let us discuss whether or not there are other methods
to estimate the values of the moments
from the form factors at low $q$, if the precise experimental data become available in future.
First, we will explore in detail the relationship between the moments and the form factors,
using the results of the previous papers\cite{am, beck, ott}.
Next, considering the observed relationship,
new methods to estimate the moments will be discussed.

Tables \ref{h_value} and \ref{he_value}
summarize the values of the $n$th moments, $\avr{r^n}_c$, of $\rho_c(r)$
obtained in the AV8'+$3N$-calculations and various analyses for $^{3}$H and $^3$He.
All the values are listed in units of fm$^n$.
The values in the AV-row
are obtained by adding the relativistic corrections
to the ones in the square brackets, according to Eqs.(\ref{2nd}) to (\ref{8th}).
In the parentheses the experimental errors are given, when available in the references.
It is seen that even the values of the msr depend on the analysis-methods,
and there are large differences between the values of the higher moments.

The values in the PA-row in the tables indicate those given
by the polynomial ansatz(PA).
It has been employed for the analyses of the low $q$ data,
as one of the `model-independent' ways in Ref.\cite{ott}.
The PA-method expands the experimental form factor, $F_{c,{\rm exp}}(q^2)$, as
\cite{ott, beck,sza}
 \begin{equation}
F_{c,{\rm exp}}(q^2)=1+\sum_{i=1}^Na_iq^{2i},\label{pa}
\end{equation}
where the value of $a_i$ is determined by the $\chi^2$-fitting to the experimental data.
The values listed in Tables \ref{h_value} and \ref{he_value}
are obtained by assuming the relationship between Eqs.(\ref{pol}) and (\ref{pa})
up to $N$ as
\footnote{The values of $a_i$ for $^3$He are
 calculated by the authors using the experimental data in Ref.\cite{ott}}

\begin{equation}
a_i=(-1)^i\frac{1}{(2i+1)!}\avr{r^{2i}}_c.\label{ass}
 \end{equation}
Most of the PA-analyses provide a good agreement with experiment, but
due to the definition\cite{ott,beck,sza}.
Ref.\cite{beck} analyzed the data for $0.0477\le q^2 \le 2.98$
fm$^{-2}$ with $N=4$. Table \ref{h_value} gives, for example, 
\begin{equation}
F_{c,{\rm PA}}(q^2=2.98)=1.026-1.429+0.933-0.309+0.040=0.262,\label{hex}
\end{equation}
in neglecting the errors.
Each value in the right-hand side corresponds to the one of Eq.(\ref{pa})
up to the eighth moment,
while the experimental value is given to be $0.282(0.007)$\cite{beck}.
For $^3$He, 
 Ref.\cite{ott} analyzed the data for $0.200\le q^2 \le 3.700$ fm$^{-2}$
 with $N=5$, and Table \ref{he_value} provides, for example,
 \begin{equation}
F_{c,{\rm PA}}(q^2=3.500)=1-2.272+3.277-3.441+2.197-0.600=0.161.\label{heex}
 \end{equation}
The obtained value reproduces well the experimental one, $0.1599(0.0031)$\cite{ott}.
Notice that the PA-method is not equivalent to the one using the identity
of Eq.(\ref{pol}) with Eq.(\ref{leib}) at small $q$. 
In Eq.(\ref{hex}), the second value, which is responsible mainly for
determination of $\avr{r^2}_c$, is over the first one which should be $1$.
In the same way, 
Eq.(\ref{heex}) does not satisfy the Leibniz criteria in Eq.(\ref{leib}).
Thus, they are not the alternating series which are
guaranteed to be convergent,
and the magnitudes of their remainder terms are uncertain,
in spite of the fact that
the set of the parameters, $a_i$, yields a good agreement with experiment.
Nevertheless, we have listed the values of the PA-moments
assuming Eq.(\ref{ass}) in Tables \ref{h_value} and \ref{he_value},
because they are useful for later discussions.

Form now on, we will discuss mainly
the form factor of $^3$He, referring to
the experimental values in Ref.\cite{ott},
because the experimental data on $^3$He
were accumulated more than on $^3$H\cite{am}, and 
Ref.\cite{sickhe4} confirmed that Ref.\cite{ott}
presented the experimental values with the lowest uncertainties,
compared with others.

The SOG-analysis\cite{am} took into account of the `world data' obtained
at that time, which are in the region of
$0.0316$\cite{sza}$\le q^2 \le 30$\cite{am} fm$^{-2}$ in $^3$He.
Fig. \ref{he_moment}-(a) shows that the integrand of the SOG-msr
is similar to the AV8'+$3N$-one,
but, as in Eqs.(\ref{pol}) to (\ref{fv2}),
the value of the msr is not determined independently
of those of the higher moments which have been shown not to be reliable
in other figures.
For example, at $q^2=1.056$ fm$^{-2}$,
the SOG-form factor is given by Table \ref{he_value} as
\begin{equation}
F_{\rm c,SOG}(q^2=1.056)\approx 1-0.67566+0.28756-0.09741+0.02690
 \approx 0.5414,\label{fv3}
 \end{equation}
corresponding to each term of Eq.(\ref{pol}) up to the eighth moment.
Ref.\cite{ott} provides the experimental value at $q^2=1.056$ fm$^{-2}$
to be $0.5359(0.0046)$. Considering the remainder term to be negative,
the value of Eq.(\ref{fv3}) is in a good agreement with experiment.
It should be noticed, however, that the value of the msr given by the second term
depends on those of the higher moments with non-negligible magnitude.

The larger the momentum transfer is, the larger the contribution of the higher moments
to the form factor is.
In $^3$He, the Leibniz criteria, Eq.(\ref{leib}), requires at least $q^2 <1.563$ fm$^{-2}$ 
for $1>q^2\avr{r^2}_c/3!$ with $\avr{r^2}_c=3.839$ fm$^{-2}$.
According to Table \ref{he_value}, the SOG eighth moment contributes to the form factor by
$78448.51\times q^8/9!=0.1095$ against the second one by
$3.839\times q^2/3!=0.9598$ at $q^2=1.5$ fm$^{-2}$,
while the AV8'+$3N$-one by $20925.242\times q^8/9!=0.2919$ 
against by $3.964\times q^2/3!=0.9910$.
The eighth moments contribute to the first digit after decimal point in the same way as the
second moments, 
but the contribution from the AV8'+$3N$-moment is larger by about $2.7$
times than that from the SOG-one.
Moreover, the values of the eighth-moment terms in both the SOG- and AV8'+$3N$-expansions
require contributions of higher moments for their convergence.
Thus, the value of the msr is not determined independently of
those of higher moments in each analysis.  


One comment should be added.
As in the previous analyses with the SOG-, FB- and PA-methods,
the values of the msr have a  possibility to be estimated,
even from experimental data in the region of high $q$ only,
because the form factor squared depends on $\avr{r^2}_c$ at any $q$. 
The cross section is proportional to $|F_c(q)|^2$, which is written as
\begin{equation}
|F_c(q)|^2=\sum_{n=0}^\infty(-1)^nq^{2n}\sum_{k=0}^n\frac{\avr{r^{2k}}_c.
\avr{r^{2(n-k)}}_c}
{(2k+1)!(2n-2k+1)!}.
\end{equation}
The $q^2$-term depends on $\avr{r^2}_c$ only, while in all the other $q$-dependent terms,
$\avr{r^2}_c$ contributes to the form factor, but accompanied
by higher moments, $\avr{r^n}_c(n\ge 4)$. 
Thus, the value of the msr may be obtained by the analysis of the high $q$-data only,
but at the same time those of the higher moments must be well known.

In order to determine the value of $\avr{r^2}_c$,
it is better to use experimental values at small $q$
for avoiding ambiguous contributions from the higher moments. 
The smallest momentum transfer in Ref.\cite{ott} is $q^2=0.200$,
which provides
the experimental value, $F_{c,{\rm exp}}(q^2=0.200)=0.8816(0.0096)$.
Against this value, Table \ref{he_value} gives
 \begin{eqnarray}
F_{c,{\rm AV}}(q^2=0.200)=1-0.13213+0.01188-0.00102+0.00009=0.8788, \label{fv4}&&     \\
F_{c,{\rm SOG}}(q^2=0.200)=1-0.12797+0.01031-0.00066+0.00003=0.8817,\label{fv5} &&     \\
F_{c,{\rm FB}}(q^2=0.200)=1-0.11650+0.00738-0.00031+0.00001=0.8906. \label{fv6}&&    \\
F_{c,{\rm PA}}(q^2=0.200)=1-0.12980+0.01070-0.00064+0.00002=0.8803,\label{fv7} &&        
\end{eqnarray}
Table \ref{he_value} lists the various values of $\avr{r^2}_c$,
but all the analyses reproduce the experimental value within error,
even without referring to the errors of each analysis.  
Hence, even the second digit after decimal point in the value of $\avr{r^2}_c$ is not
fixed by these comparisons.
As far as the moments are concerned, it is not clear whether or not the SOG-method
is better than others, even if the world data are taken account of. 

When we compare the results of various analyses in Eqs.(\ref{fv4}) to (\ref{fv7})
with the experimental value of the $10^{-2}$-accuracy as $0.8816(0.0096)$ in Ref.\cite{ott},
we understand first that
it is necessary to improve the accuracy of experiment as less than $10^{-3}$,
in order to obtain the more
accurate values of $\avr{r^2}_c$. 
Second, at the same time, the experiments should be performed in a smaller $q$ region 
of $0< q^2 \le 0.200$ fm$^{-2}$, so as to make contributions from the higher moments
as small as possible.
Recently, low $q$-experiments are actually in progress at $0.0025 \le q^2 \le 0.065$
fm$^{-2}$ on $^{208}$Pb with accuracy of the cross sections with $10^{-3}$
(T. Suda, personal communication).
If they become available on $^3$He,
there are several ways to derive not only the value of the msr,
but also those of a few low moments.
In order to discuss those methods which do not rely on the SOG- and FB-ones, 
we need to explore the above second requirement in more detail,
by using the results of the AV8'+$3N$-calculations, as a guide.

In employing the values of moments of the AV8'+$3N$-calculations which are
largest among those in Table \ref{he_value}, we have
for the first five terms of Eq.({\ref{pol}}) giving the form factor with accuracy up to $10^{-5}$,
\begin{eqnarray}
F_{c,{\rm AV}}(q^2=0.100)
 &=&1-0.066067+0.002970-0.000127+5.8\times 10^{-6}=0.93678, \label{fva}\,\,\,\,\,\,\\
F_{c,{\rm AV}}(q^2=0.050)
 &=&1-0.033033+0.000743-0.000016+3.6\times 10^{-7}=0.96769, \label{fvb}\\
F_{c,{\rm AV}}(q^2=0.005)
 &=&1-0.003303+7.4\times 10^{-6}-1.6\times 10^{-8}+\,\cdots\cdots\,\,=0.99670. \label{fvc}
\end{eqnarray}

The above equations, which are given for arbitrary values of the low $q^2$,
satisfy apparently the Leibniz criteria, Eq.(\ref{leib}),
so that we can expect from the estimation theorem, Eq.(\ref{les}), 
for example, for the sum up to the sixth moment($k=3$) in Eq.(\ref{fvb})
\begin{equation}
|F_{c,{\rm AV},k=3}(q^2=0.050)-F_{c,{\rm AV}}(q^2=0.050)| < 3.6\times 10^{-7}.\label{fvd}
\end{equation}
On the basis of the assumption that the form factor at low $q$ can be written
as the alternating series to satisfy the Leibniz criteria, as above equations,
it is possible to analyze the experimental data with following ways.

Eqs.(\ref{fva}) to (\ref{fvc}) imply that
once experiment  determines the value of $\avr{r^2}_c$ acculately at small $q$, as in Eq.(\ref{fvc}),  
then the values of higher moments will be determined sequentially with increasing $q$.

When the values of $q^2$ are not small enough to determine the value of the msr
independently of those of higher moments within the required accuracy,
the PA-method in Eq.(\ref{pa}) for the low $q$-data can be used,
considering the Leibniz criteria, Eq.(\ref{leib}).
Suppose that
experiments are performed at $q^2_{\rm min} \le q^2 \le q^2_{\rm max}$.
In the first step,
the small number of the parameters, $N$, for around the $q^2_{\rm min}$-data, is  fixed
by the estimation theorem, Eq.(\ref{les}). 
On the right-hand side of Eq.(\ref{les}),
the value of $m=N+1$ is given by
the well established calculations, for example, with AV8'+$3N$, as a guide,
according to the required significant digits, as in Eq.(\ref{fvd}).  
Confirming that the series obtained by the $\chi^2$-fitting satisfies
the Leibniz criteria, Eq(\ref{leib}), then, Eq.(\ref{ass}) is used for estimating
the values of the moments.
In this first step, the accuracy of the last $a_i$s, for example, $a_N$, may be poor.
In the next step, fixing the values of a few first $a_i$ obtained in the first step,
and increasing the values of $q$ and $N$, the PA-analysis is repeated for improving
the value of the last $a_i$ in the first step. These steps are repeated up to $q^2_{\rm max}$.

Analysis of the low $q$ experiments receives
an advantage to estimate the values of moments.
Ref.\cite{am} were interested in magnetic form factors also,
accompanied with elastic scattering.
In order to discuss electric and magnetic form factors calculated in the PWBA-framework,
they inferred the PWBA form factors from the SOG charge and magnetization densities.
Those densities were derived by the phase-shift analysis for charge scattering and
by the DWBA-analysis for magnetic scattering.
The contribution of the magnetic form factor to the cross section is reduced by the factor,
$q^2/M^2$, compared with that of the electric form factor.
In our case, the value of $q$ is small as, for example,
$q^2=0.005$ fm$^{-2}$, which yields the reduction factor
about $2.20\times 10^{-4}$.
In experiments at forward angles, the magnetic contributions are further reduced
by the factor $\tan^2\theta/2$.
Analysis of the small $q$-data is expected to be simpler than that of higher $q$.

In following the way of Ref.\cite{am} for estimation of the distortion effects on the
PWBA form factors in the above proposed method, 
the SOG-density should not be used as an input charge density in the phase-shift analyses.
At present there are reliable $\rho_{c,{\rm Theo}}(r)$s,
like that by AV8'+$3N$, for a few-body systems \cite{din}.
By using those densities for the PWBA-analyses also, the Coulomb distortion effects
would be examined, in comparing them
with ones by the various assumptions in Refs.\cite{am,beck,ott,sza}.


\section{Summary}  \label{sum}

\begin{table}
 \hspace*{1cm}%
\begin{tabular}{|l|c|c|c|c||l|c|c|c|c|} \hline
$^3$H&
$\avr{r^2}_c$&
$\avr{r^4}_c$&
$\avr{r^6}_c$&
$\avr{r^8}_c$&
 $^3$He&
$\avr{r^2}_c$&
$\avr{r^4}_c$&
$\avr{r^6}_c$&
$\avr{r^8}_c$
\\ \hline
\rule{0pt}{12pt}%
SOG  & $0.979$ &$0.900$ &$0.728$ &$0.485$&SOG &$0.977$&$0.879$&$0.660$&$0.380$ \\
FB  &  $0.892$ &$0.651$ &$0.357$ &$0.141$&FB  &$0.889$&$0.629$&$0.311$&$0.100$   \\ \hline
\end{tabular}
\caption{
The ratio of the SOG- and FB-values of the $n$th moments to the AV8'+$3N$-ones
in $^3$H and $^3$He.
The AV8'+$3N$-values do not take account of small relativistic corrections.
 For the details, see the text. 
}
 \label{ratio}
\end{table}

The sum-of-Gaussians(SOG)\cite{sog}- and the Fourier-Bessel(FB)\cite{fb}-method
to analyze the electron-scattering cross sections
have widely been used as a `model-independent' way\cite{vries},
because they need not assume a specific function like Fermi-type
for the charge density distribution, $\rho_c(r)$,
which is required as an input in the phase-shift analysis of experimental data.
It is not necessary for the `model-independent' analyses, however, to determine uniquely
$\rho_c(r)$, because experimental data available at present are
limited in a region of the momentum transfer, $q$, as $q^2_{\rm min} \le q^2 \le q^2_{\rm max}$.
As a result, the values of the moment, $\avr{r^n}_c=\int d^3r\,r^{n}\rho_c(r)/Z$,
depend on the analysis-method to deduce $\rho_c(r)$ from the experimental cross-sections.  
In the present paper, we have investigated how the values of the moments depend on 
the SOG- and FB-method in $^3$H and $^3$He.

It has been shown
that because the SOG- and the FB-method do not describe well the tails of $\rho_c(r)$,
in a quantum mechanical point of view,
the both analyses provide smaller values
 of $\avr{r^n}_c$ than those by detailed calculations with AV8'+$3N$,
 as shown in Table \ref{ratio}.
  It shows the ratios of the values of the $n$th moments derived from
 the SOG(FB)-analyses to those of the AV8'+$3N$-calculations
 without small relativistic corrections listed in Tables \ref{h_value} and \ref{he_value}.
 The disagreement between those of the AV8'+$3N$- and SOG(FB)-analyses
 increases with increasing $n$.
 In the eighth moments, the SOG-analysis provides
 less than $50\%$ of the AV8'+$3N$-values, while the FB-analysis less than $15\%$.
The values of the msr(the second moment) seem to be obtained rather well
in both analyses, compared with the AV8'+$3N$-ones, but they are not determined independently of
the disagreement of the higher moments.
The SOG(FB) form factors to reproduce the experimental ones are functions of not only  
the value of the msr, but also the values of the higher moments.
These results are not affected by the contributions of the meson-exchange currents(MEC)
to the cross sections, because the MEC-contributions appear around the center of the nucleus.
In order to determine the tails more precisely,
the `world experimental data' available at present
($0.2 \le q^2 \le 30$ fm$^{-2}$\cite{am})\footnote{Ref.\cite{sza} cited in Ref.\cite{am}
reported  the data of $F_c^2(q)$
between $0.0319\le q^2 \le 0.3362$ fm$^{-2}$, but with $1$ significant digit.}
are not enough for the SOG- and FB-methods.

More data may help to improve the SOG- and FB- analyses.
Because the tails are sensitive to the cross sections at small $q^2<0.1$ fm$^2$,
however, other methods to determine the values of moments are proposed,
without relying on the `world data'\cite{am}.
They are based on the polynomial ansatz(PA) in Ref.\cite{ott}, but
under the constraint on the alternating series according to the Leibniz criteria.
The new methods require the accurate experiments providing form factors
with the accuracy less than
$10^{-3}$ in a region, $0 < q^2 \le 0.1$ fm$^{-2}$.

If values of the msr are determined accurately by the new method,
comparing them with the values from the muonic atom experiment\cite{atom},
those of the higher moments may be estimated at the same time more accurately
than those listed in Table \ref{ratio}. 
Then, the neutron distributions are explored on the same basis as the proton ones\cite{ks1}. 
As a result, the understanding of the mirror nuclei is expected to be extremely deepened.
The accurate determination of the various moments
thus makes it possible for a few-body systems
to play a role as reference nuclei for investigation of
more complex nuclei\cite{din, sickhe4,hagen}.
Determination of the higher moments may also be useful for discussions of
Zemach moments\cite{din,sickz}. 
Experiments with high accuracy at low $q$-region are desired.

\section*{Acknowledgments}
The authors would like to thank Professor H. Kurasawa,
Professor K. Tsukada  and Professor T. Suda
for useful discussions.
Most of the numerical calculations on the electron scattering
are indebted to Professor Kurasawa and on the PA-moments 
to Professor K. Tsukada.
This work was supported by JSPS KAKENHI Grant Numbers JP22K18706
and partially by ERATO project-JPMER2304, JPS KAKENHI grant Number 23K03378.

\theendnotes

\end{document}